\documentclass[12pt]{article}
\usepackage{epsfig, amsmath, amssymb}
\usepackage{graphicx,psfrag}
\usepackage{xcolor}
\usepackage{physics}
\usepackage{mathrsfs}
\newcommand{\nn}{\nonumber}
\newcommand{\be}{\begin{equation}}
\newcommand{\ee}{\end{equation}}
\newcommand{\ben}{\begin{equation}}
\newcommand{\een}{\end{equation}}
\newcommand{\bea}{\begin{eqnarray}}
\newcommand{\eea}{\end{eqnarray}}
\newcommand{\bA}{\begin{array}}
\newcommand{\eA}{\end{array}}
\newcommand{\bc}{\begin{center}}
\newcommand{\ec}{\end{center}}
\newcommand{\al}{\alpha}

\newcommand{\ra}{\rightarrow}
\newcommand{\del}{\partial}

\newcommand{\ie}{{\it i.e.}}
\newcommand{\eg}{{\it e.g.}}

\def\BZ{{\mathbb Z}}

\newcommand{\lan}{\langle}
\newcommand{\ran}{\rangle}

\numberwithin{equation}{section}

\begin{document}


\begin{titlepage}

%

\bc

\hfill 
\\         [25mm]

{\Huge Notes on time entanglement 
  \\ [2mm] and pseudo-entropy}

\vspace{16mm}

{\large K.~Narayan,\ \ Hitesh K. Saini} \\
\vspace{3mm}
{\small \it Chennai Mathematical Institute, \\}
{\small \it H1 SIPCOT IT Park, Siruseri 603103, India.\\}

\ec
\vspace{30mm}

\begin{abstract}
  Following arXiv:2210.12963 [hep-th], we investigate aspects of the time
  evolution operator regarded as a density operator and associated
  entanglement-like structures in various quantum systems.  These
  involve timelike separations and generically lead to complex-valued
  entropy, although there are interesting real subfamilies. There are
  many parallels and close relations with reduced transition matrices
  and pseudo-entropy, which we discuss and clarify.  For instance, a
  related quantity involves the time evolution operator along with a
  projection onto some initial state, which amounts to analysing
  pseudo-entropy for the initial state and its time-evolved final
  state.
\end{abstract}


\end{titlepage}

{\tiny 
\begin{tableofcontents}
\end{tableofcontents}
}


\vspace{-4mm}

\section{Introduction}

Generalizations of the Ryu-Takayanagi formulation of holographic
entanglement \cite{Ryu:2006bv,Ryu:2006ef,HRT} in $AdS/CFT$
\cite{Maldacena:1997re,Gubser:1998bc,Witten:1998qj} to de Sitter space
reveal new fascinating structures. These are based on taking the
future boundary $I^+$ of de Sitter space as the anchoring surface for
extremal surfaces, along the lines of $dS/CFT$
\cite{Strominger:2001pn,Witten:2001kn,Maldacena:2002vr,Anninos:2011ui}.
Most recently these appear in \cite{Doi:2022iyj,Narayan:2022afv},
refining previous investigations of extremal surfaces and holographic
entanglement in de Sitter space
\cite{Narayan:2015vda,Narayan:2015oka,Sato:2015tta,Miyaji:2015yva,
  Narayan:2017xca,Narayan:2020nsc,Hikida:2022ltr,Hikida:2021ese} (see
also \cite{Arias:2019pzy,Cotler:2023xku}).

In the present work, we explore aspects of
``time-entanglement'', or timelike entanglement, in various quantum
mechanical systems, towards understanding entanglement-like structures
involving timelike-separations, following
\cite{Narayan:2022afv}. There are close parallels with pseudo-entropy
\cite{Nakata:2020luh} (and \cite{Doi:2022iyj}), as we will
describe. Related investigations appear in
\eg\ \cite{Mollabashi:2020yie}-\cite{Chen:2023gnh} (also
\cite{Wang:2018jva}).

To summarize the de Sitter studies (from \cite{Narayan:2022afv}),
extremal surfaces anchored at $I^+$ turn out to not return to $I^+$
(unlike those in $AdS$ where the surfaces possess turning
points). Since such surfaces do not return, they require extra data or
boundary conditions in the past (interior). In entirely Lorentzian de
Sitter spacetime, this leads to future-past timelike surfaces
stretching between $I^\pm$. Apart from an overall $-i$ factor
(relative to spacelike surfaces in $AdS$) their areas are real and
positive. With a no-boundary type boundary condition, the top half of
these timelike surfaces joins with a spacelike part on the hemisphere
giving a complex-valued area.  Since these surfaces necessarily have a
timelike component (or run along a complex time contour), they have
complex areas.
Two aspects of ``time-entanglement'' in simple toy models in quantum
mechanics were described in \cite{Narayan:2022afv}. One is based on a
future-past thermofield double type state entangling timelike
separated states, which leads to entirely positive structures.
Another is based on the time evolution operator and reduced transition
amplitudes, which leads to complex-valued entropy.


In the present paper we discuss various aspects of the time evolution
operator regarded as a density operator and its entanglement
structures which involve timelike separations. There are many
parallels and close relations with pseudo-entropy
\cite{Nakata:2020luh}: we summarize some central points on time
evolution and pseudo-entropy in sec.~\ref{sec:summary}, including a
general map in sec.~\ref{sec:tEp-TM}. We then study various classes of
finite quantum mechanical examples in sec.~\ref{sec:tE-examples},
including qubit systems and harmonic oscillators (some detailed in
Appendices~\ref{app:tE-pE}, ~\ref{app:qubitChains} and
~\ref{app:cpldOsc}). In sec.~\ref{sec:tE-proj} we study entanglement
structures for the time evolution operator along with a projection
operator onto some state towards isolating components of the time
evolution operator. This ends up amounting to pseudo-entropy for this
state and its time-evolution: in sec.~\ref{sec:tE-proj-TFD} we study
thermofield-double type states and find that some general features
emerge. In sec.~\ref{sec:tE-normt=0} we study the time evolution
operator normalized at $t=0$ (rather than at general time $t$): this
gives rise to various detailed differences in the entanglement
structures that emerge.  In sec.~\ref{sec:2dCFT-tE} we describe some
aspects of entanglement entropy in 2-dim CFT for timelike intervals,
elaborating on
that in \cite{Narayan:2022afv}. Some of the discussions here have
partial overlap with \cite{Doi:2022iyj,Doi:2023zaf}. In these
time-independent situations so far, the structure of time-entanglement
shows parallels with ordinary finite temperature entanglement,
but with analytic continuation to imaginary temperature
$\beta=it$. In sec.~\ref{sec:t-depInt}, we study time-dependent
interactions focussing on simple 2-qubit systems with $\delta$-function
potentials, and the resulting time entanglement.

Overall, pseudo-entropy \cite{Nakata:2020luh} is a generalization of
entanglement entropy involving two arbitrary states (without
necessarily specifying dynamical information): this does not need to
pertain to timelike separations per se. The notions of time
entanglement are designed to deal with timelike separations, involving
entanglement structures based on the time evolution operator, as well
as projection onto specific initial states: so in particular we
require specifying a Hamiltonian that dictates time evolution. However
the calculations involved in studying time entanglement entropy are
closely related to those in evaluating pseudo-entropy
\cite{Nakata:2020luh}. Our goal in these notes is more an exploration
of time entanglement and how it dovetails with pseudo-entropy, rather
than a detailed classification (which already appears for
pseudo-entropy of various quantum systems in \cite{Nakata:2020luh}
and subsequent work).

\section{Summary: time evolution and pseudo-entropy}\label{sec:summary}

Our investigations, following \cite{Narayan:2022afv}, are based on
regarding the time evolution operator as a density operator,
performing partial traces over subsystems and evaluating the
corresponding von Neumann entropy. The time evolution operator
${\cal U}(t)=e^{-iHt}$ for a system with Hamiltonian $H$ can be written
in terms of (time-independent) Hamiltonian eigenstates $|i\ran$ (which
are defined on some past time slice $P$).
Then the time evolution operator normalized at an arbitrary time $t$
gives
\bea\label{rhot(t)2}
\quad {\cal U}(t) = e^{-iHt} = \sum_i e^{-iE_it}|i\ran\lan i| \!
&=& \! \sum_i |i\ran_t\,\lan i|_{_P}\,,\qquad\quad
|i(t)\ran\equiv |i\ran_t=e^{-iE_it}|i\ran_{_P}\,;\nn\\
\rho_t(t) \equiv {{\cal U}(t)\over {\rm Tr}\,{\cal U}(t)}
\quad &\Rightarrow&\quad
\rho_t(t) = \sum_i p_i\, |i\ran_{_P}\lan i|_{_P}\, ,
\qquad p_i={e^{-iE_it}\over \sum_j e^{-iE_jt}}\ , \nn\\
\ra\ \ \
\rho_t^A = {\rm Tr}_{_B} \rho_t = \sum_{i} p_i'\, |i'\ran_{_P}\lan i'|_{_P}
\quad &\ra&\quad
S_A = -\sum_i p_i'\,\log p_i'\ .
\eea
As is clear, there are sharp parallels with ordinary finite temperature
entanglement structures, except with imaginary temperature $\beta=it$\,:
this will be seen explicitly as a recurring theme throughout much of
what follows.\\
A related quantity involves the time evolution operator with
projection onto some state $|i\ran$,
\be\label{tE-pr}
\rho_t^{|i\ran}
= {\rho_t|i\ran\lan i|\over {\rm Tr} (\rho_t|i\ran\lan i|)}
= {|f[i](t)\ran \lan i|\over {\rm Tr}(|f[i](t)\ran \lan i|)}\,,\qquad
|f[i](t)\ran = e^{-iHt}|i\ran\,;\qquad
\rho_t^{|i\ran, A}={\rm Tr}_{_B}\,\rho_t^{|i\ran}\ .
\ee
The state $|f[i]\ran$ is the final state obtained by time-evolving
the initial state $|i\ran$.\ We obtain
\be
|i\ran = \sum c_n|n\ran\ ;\qquad
\rho_t^{|i\ran} = {1\over \sum_{k} e^{-iE_kt}|c_k|^2}
\sum_{k,m} e^{-iE_kt}c_kc_m^*|k\ran\lan m|\ 
\ee
for a general (non-eigen)state $|i\ran$.
At $t=0$, the time evolution operator is just the identity operator,
a sum over all the eigenstate projection operators, while the time
evolution operator with projection becomes simply the density matrix
for the initial state $|i\ran$.
For any nonzero time $t$, there is timelike separation between
the initial states $|\psi\ran_{_P}$ and the eventual states $|\psi\ran_t$.
These entanglement structures involving timelike separations and time
evolution have close parallels with pseudo-entropy \cite{Nakata:2020luh}
obtained from the reduced transition matrix for two arbitrary states
$|i\ran, |f\ran$\,:
\be\label{tEpE}
{\cal T}_{f|i}^A = {\rm Tr}_B
   \left({|f\ran\lan i|\over {\rm Tr}(|f\ran\lan i|)} \right)\,.
\ee

To summarise in generality, consider a bipartite system, the Hilbert
space being characterized by Hamiltonian eigenstates $|i,i'\ran$
with energies $E_{i,i'}$.
The normalized time evolution operator (\ref{rhot(t)2}) and its
partial trace over $B\equiv \{i'\}$ are
\be\label{tE1-tE}
\rho_t = {1\over\sum_{i,i'} e^{-iE_{i,i'}t}}
\sum_{i,i'} e^{-iE_{i,i'}t}\,|i,i'\ran\lan i,i'|\quad\ra\quad
\rho_t^A = {1\over\sum_{i,i'} e^{-iE_{i,i'}t}}
\big(\sum_{i'} e^{-iE_{i,i'}t}\big)\,|i\ran\lan i|\ .
\ee
The time evolution operator with projection onto state $|I\ran$ is
\bea\label{tE2-tEp}
&& |I\ran=\sum_{k,k'}c_{k,k'}|k,k'\ran\,,\qquad
\rho_t^{|I\ran} = {1\over \sum_{i,i'}|c_{i,i'}|^2e^{-iE_{i,i'}t}}
\sum_{i,i',j,j'} c_{i,i'}c_{j,j'}^*e^{-iE_{i,i'}t}|i,i'\ran\lan j,j'|\,, \nn\\
&& \qquad\qquad
\rho_t^{|I\ran, A} = {1\over \sum_{i,i'}|c_{i,i'}|^2e^{-iE_{i,i'}t}}
\sum_{i,j} \big(\sum_{i'} c_{i,i'}c_{j,i'}^*e^{-iE_{i,i'}t}\big)|i\ran\lan j|\ .
\eea
The reduced transition matrix for pseudo-entropy is obtained as
\bea\label{tE3-pE}
&& |I\ran = c_{i,i'}|i,i'\ran ,\quad |F\ran = c_{i,i'}'|i,i'\ran\,;\qquad
{\cal T}_{F|I} = {1\over \sum_{i,i'} c_{i,i'}'c_{i,i'}^*}\,
\sum_{i,i',j,j'} c_{i,i'}' c_{j,j'}^*\,|i,i'\ran\lan j,j'|\ ,\nn\\
&& \qquad\qquad {\cal T}_{F|I}^A = {1\over \sum_{i,i'} c_{i,i'}'c_{i,i'}^*}\,
\big(\sum_{i'} c_{i,i'}' c_{j,i'}^*\big)\,|i\ran\lan j|\ .
\eea
It is clear that the time evolution operator with projection
(\ref{tE2-tEp}) is obtained from the pseudo-entropy reduced transition
matrix (\ref{tE3-pE}) by restricting to the final state being that
obtained by time-evolving the initial state, \ie\
$|F\ran={\cal U}(t)|I\ran$.

\subsection{The time evolution operator and the 
  transition matrix}\label{sec:tEp-TM}

With a single Hilbert space, the structure of the reduced transition
matrix appears different in detail from that of the reduced time
evolution operator: this is clear in bipartite systems from
(\ref{tE1-tE}), (\ref{tE2-tEp}), (\ref{tE3-pE}).
However it would seem that there should be close connections between
the time evolution operator and the transition matrix since both
pertain to time evolution if we focus on final states as time-evolved
initial states.

Towards studying this, let us first recall that a special class of
states comprises thermofield-double type states\
$|I\ran_{TFD} = \sum_k c_{k,\{k\}}|k,\{k\}\ran$,\ with only diagonal
components (a further special subclass comprises maximally entangled
TFD states, with all $c_{k,\{k\}}$ equal).

Towards mapping time evolution and the transition matrix, consider
doubling the Hilbert space at both initial and final times:
\ie\ extend the Hilbert state $\mathscr{H}\equiv\mathscr{H}_1$ to
$\mathscr{H}_1 \otimes \mathscr{H}_2$, where the Hilbert space
$\mathscr{H}_2$ is an identical copy of $\mathscr{H}_1$. Now consider
thermofield-double type initial and final states:
\be
|\psi_I\ran = \sum_i c_i^I\, |i\ran_1|i\ran_2\,,\qquad
|\psi_F\ran = \sum_i c_i^F\, |i\ran_1|i\ran_2\,,
\ee
where $\{|i\ran\}$ is a basis of states.\
The (un-normalized) transition matrix is
\be
{\cal T}_{F|I} = |\psi_F\ran\lan \psi_I|
= \sum_{i,j} c_i^F c_j^{I\,*}\ |i\ran_1|i\ran_2\ \lan j|_1\lan j|_2\,.
\ee
Performing a partial trace over copy-2 gives
\be
{\rm Tr}_2\,{\cal T}_{F|I} = \sum_i c_i^F c_i^{I\,*}\ |i\ran_1\lan i|_1\ .
\ee
For this to equal the time evolution operator, we require
\be\label{TF|I->U(t)}
{\rm Tr}_2\,{\cal T}_{F|I} = 
{\cal U}(t) = \sum_i e^{-iE_it}\, |i\ran\lan i|\qquad\Rightarrow\qquad
c_i^F c_i^{I\,*} = e^{-iE_it}\ .
\ee
A ``symmetric'' solution is
\bea\label{TF|I->U(t)2}
&& c_i^I = e^{iE_it/2}\,: \qquad
|\psi_I\ran = \sum_i e^{iE_it/2}\, |i\ran_1|i\ran_2\,,  \nn\\
&& c_i^F = e^{-iE_it/2}\,: \qquad
|\psi_F\ran = \sum_i e^{-iE_it/2}\, |i\ran_1|i\ran_2\,.
\eea
These can be regarded as obtained from a continuation $\beta\ra it$ of
the usual finite temperature thermofield-double type states\
$e^{-\beta E_i/2}|i\ran|i\ran$.  There are of course less symmetric
solutions $c_i^I,\ c_i^F$, describing the initial and final states.
However the symmetric solution reduces to ordinary entanglement when
the initial and final states are the same, \ie\ 
$|\psi_I\ran=|\psi_F\ran$\ (\ie\ at $t=0$), the transition matrix becomes
the usual density matrix\ ${\cal T}_{F|I}=|\psi_I\ran\lan \psi_I|=\rho_I$\
for the state $|\psi_I\ran$. Thus the time evolution operator can
be regarded as a particular reorganization of the transition matrix
appearing in pseudo-entropy.

It is worth noting that for systems with infinite towers of states,
the trace of the time evolution operator contains highly oscillatory
terms and thus requires a regulator to be well-defined: we will see
this explicitly for the harmonic oscillator later; see
(\ref{tE-HO-reg}).

\medskip

{\bf Single qubit:}\ \ This simple case serves to illustrate the
above. In this case (described by (\ref{qm2state})), we have\
$H\ket{1}=E_{1}\ket{1},\ H\ket{2}=E_{2}\ket{2}$, with $H$ the
Hamiltonian. Let us take
\be
\ket{\psi_F} = \sum_{n=1,2} e^{- \frac{ i\, E_n t}{2}} \ket{n}_1 \otimes \ket{n}_2 \,, \qquad
\ket{\psi_I} = \sum_{m=1,2} e^{ \frac{ i\, E_m t}{2}} \ket{m}_1 \otimes \ket{m}_2 \,.
\ee
Here the subscript 2 stands for the second auxiliary system with the
identical Hilbert space $\mathscr{H}_2$.
Then the unnormalised transition matrix\
$T= \ket{\psi_F} \bra{\psi_I}$\ is
\be
T_{F|I} = \ket{\psi_F} \bra{\psi_I} =  \sum_{n,m=1,2} \, e^{\frac{-i(E_n+E_m)t}{2}}
\ket{n}_1 \ket{n}_2 \bra{m}_1 \bra{m}_2 \,.
\ee
Taking a partial trace over the second component gives
\be
T^1_{F|I} = Tr_2(T_{F|I}) = \sum_{n=1,2} \, e^{-i\, E_n t } \ket{n}_1\bra{n}_1 \,
=e^{-iHt}  \,, 
\ee
thus obtaining the time evolution operator. This illustrates the
general discussion earlier in this simple case.


\section{Time evolution operator \& entanglement:\ examples}
\label{sec:tE-examples}

In this section we will study various examples of finite quantum systems
to explore the entanglement structure of the time evolution operator.

\subsection{2-qubit systems}

For a 2-state system,
\be\label{qm2state}
H|k\ran = E_k|k\ran\,, \quad k=1,2\ ;\qquad\quad
|k\ran_F \equiv |k(t)\ran = e^{-iE_kt}|k\ran_P\ . \qquad\quad [\lan 1|2\ran=0]
\ee
we obtain\ $\rho_t(t)$ using (\ref{rhot(t)2}). Now, imagining a 2-spin
analogy $|1\ran\equiv |++\ran,\ |2\ran\equiv |--\ran$, performing a
partial trace over the second spins gives
\bea\label{rhotA-2state}
&& \rho_t^A = {1\over 1+e^{i\theta}} \big( |+\ran_P\lan +|_P
+ e^{i\theta} |-\ran_P\lan -|_P \big) ,\qquad\quad \theta=-(E_2-E_1)t\,,
\nn\\ [1mm]
&& S_A = -{\rm tr} \big(\rho_t^A\log\rho_t^A)
= -{1\over 1+e^{i\theta}} \log {1\over 1+e^{i\theta}}
- {1\over 1+e^{-i\theta}} \log {1\over 1+e^{-i\theta}} \,,
\eea
so the von Neumann entropy, recast as $\al+\al^*$, is real-valued
in this special case. We see that $S_t^A$ grows large as
$\theta\ra (2n+1)\pi$. Further $\rho_t^A$ and $S_t^A$ are periodic in
$\theta$ and so in time $t$\ (simplifying $S_t^A$ shows terms
containing $\log (e^{i\theta/2})$ which we retain as it is, rather
than ${i\theta\over 2}$\,, so as to avoid picking specific branches
of the logarithm, thereby losing manifest periodicity; within one
$\theta$-cell the simplified expression for $S_t^A$ coincides with
the corresponding one  in \cite{Nakata:2020luh}).

Now consider two qubits, each being $|1\ran, |2\ran$, with a more
general Hamiltonian
\be\label{H-2qubit}
H = E_{11}|11\ran\lan 11| + E_{22}|22\ran\lan 22|
+ E_{12} \big( |12\ran\lan 12| + |21\ran\lan 21| \big)
\ee
that is diagonal in this basis. It is reasonable to take $E_{12}=E_{21}$.
So the normalized time evolution operator (\ref{rhot(t)2}) becomes
\bea\label{rhot-2qubit}
&& \rho_t = \sum_{i,j}
          {e^{-iE_{ij}t}\over \sum_{kl} e^{-iE_{kl}t}}\, |ij\ran\lan ij| 
\,=\, {\big(|11\ran\lan 11|
+ e^{i\theta_1}|22\ran\lan 22| + e^{i\theta_2} ( |12\ran\lan 12|
+ |21\ran\lan 21| )\big) \over 1+e^{i\theta_1}+2e^{i\theta_2}}\,; \nn\\ [2mm]
&&\qquad\qquad\qquad\qquad
\theta_1\equiv -(E_{22}-E_{11})t\,,\qquad \theta_2\equiv -(E_{12}-E_{11})t\,.\
\eea
(At $t=0$, the $\theta_i$ vanish and this is the normalized identity
operator.)
A partial trace over the 2nd component gives the reduced time evolution
operator,
\be\label{rhotA-2qubit}
\rho_t^A = {1\over 1+e^{i\theta_1}+2e^{i\theta_2}} \Big( \big(1+e^{i\theta_2}\big)
|1\ran\lan 1| + \big(e^{i\theta_1}+e^{i\theta_2}\big) |2\ran\lan 2| \Big)
\ee
which generically has complex-valued von Neumann entropy. It is
clear that this matches ordinary finite temperature entanglement,
except with imaginary temperature $\beta=it$.

Now let us impose an exchange symmetry $|1\ran\leftrightarrow|2\ran$\,:
this occurs for instance if we consider two spins $|\pm\ran$ with
nearest neighbour interaction $H=-Js_z^1s_z^2$. This restriction now
implies $E_{22}=E_{11}$ thereby reducing (\ref{rhotA-2qubit}) to
(\ref{rhotA-2state}) earlier, with just one nontrivial phase,
giving real entropy.\\
{\bf Qubit chains:}\ \ In Appendix~\ref{app:qubitChains}, we study
finite and infinite chains of qubits with nearest neighbour
interactions, towards understanding the reduced time evolution operator
for a single qubit, after partial trace over all other qubits. This
also reveals interesting complex-valued entropy in general, obtainable
as a finite temperature system but with imaginary temperature. We
also find a real-valued slice when the system enjoys 
$|1\ran\leftrightarrow|2\ran$ exchange symmetry.

To illustrate obtaining the time evolution operator (\ref{rhot-2qubit})
from the doubled transition matrix as in (\ref{TF|I->U(t)}),
(\ref{TF|I->U(t)2}), we write
\be
\ket{\psi_F} = \sum_{n,m=1,2}  e^{- \frac{ i\, E_{nm} t}{2}} \ket{nm}_1 \otimes \ket{nm}_2 \,,\qquad
\ket{\psi_I} = \sum_{n,m=1,2}  e^{ \frac{ i\, E_{nm} t}{2}}  \ket{nm}_1 \otimes \ket{nm}_2 \,.
\ee
Then the unnormalized transition matrix $T= \ket{\psi_F} \bra{\psi_I}$
after partial trace over the second component gives
\begin{align}
T^1_{F|I} = {\rm Tr}_2\Big( \sum_{n,m,p,q=1,2} \, e^{- \frac{ i E_{nm} t}{2}} e^{- \frac{ i E_{pq} t}{2}} \ket{nm}_1 \ket{nm}_2 \bra{pq}_1 \bra{pq}_2 \Big)
= \sum_{n,m=1,2} \, e^{-i\, E_{nm} t } \ket{nm}_1\bra{nm}_1\,,
\end{align}   
so this reduced transition matrix is the same as the unnormalized
time evolution operator.

\subsubsection{Mutual information}

Mutual information defined as\ $I[A,B] = S[A]+S[B]-S[A\cup B]$ can be
studied for the time evolution operator as well. In the general 2-qubit
case (\ref{H-2qubit}), (\ref{rhot-2qubit}), above, we can calculate
$\rho_t^1={\rm Tr}_2\rho_t$ and $\rho_t^2={\rm Tr}_1\rho_t$, which then
leads to the von Neumann entropies $S_t^1$ and $S_t^2$ respectively.
The time evolution operator $\rho_t$ itself leads to
$S_t=-{\rm tr} \big(\rho_t\log\rho_t)$.\ It is straightforward to
see that $\rho_t^{1,2}$ are of the same form as $\rho_t^A$ in
(\ref{rhotA-2qubit}), which alongwith $\rho_t$ in (\ref{rhot-2qubit})
gives
\bea
&& S_t^{1,2} = -{1+e^{i\theta_2}\over 1+e^{i\theta_1}+2e^{i\theta_2}}\,
\log {1+e^{i\theta_2}\over 1+e^{i\theta_1}+2e^{i\theta_2}}\, -\,
{e^{i\theta_1}+e^{i\theta_2}\over 1+e^{i\theta_1}+2e^{i\theta_2}}\,
\log {e^{i\theta_1}+e^{i\theta_2}\over 1+e^{i\theta_1}+2e^{i\theta_2}}\ , \nn\\
&& S_t = -{1\over 1+e^{i\theta_1}+2e^{i\theta_2}}\,
\log {1\over 1+e^{i\theta_1}+2e^{i\theta_2}}\,
-\,{e^{i\theta_1}\over 1+e^{i\theta_1}+2e^{i\theta_2}}\,
\log {e^{i\theta_1}\over 1+e^{i\theta_1}+2e^{i\theta_2}}\, \qquad \nn\\
&& \qquad\qquad\qquad\ -\, {2e^{i\theta_2}\over 1+e^{i\theta_1}+2e^{i\theta_2}}\,
\log {e^{i\theta_2}\over 1+e^{i\theta_1}+2e^{i\theta_2}}\,, 
\eea
so the mutual information is
\be
I[A,B] = S_t^1+S_t^2-S_t\ .
\ee
In general this is nonzero and complex since the entropies are complex
in general.
However there are special cases: for instance if all energy eigenvalues
are identical, then
\be
\theta_{1,2}=0\,:\qquad S_t^{1,2}=\log 2\,,\ \ S_t=2\log 2
\quad\Rightarrow\quad  I[A,B]=0\ ,
\ee
although the time evolution is nontrivial since each phase $e^{-iEt}$
is nonzero.\\
Likewise the 2-state subcase (\ref{qm2state}) is obtained by setting
$e^{i\theta_2}=0$ which gives $S_t^{1,2}, S_t$ of the same real-valued
form as in (\ref{rhotA-2state}), so $I[A,B]=S_t^1$.

These expressions above can also be viewed as arising from the finite
temperature results for inverse temperature $\beta$ continued to
$\beta=it$. From that point of view, the high temperature limit
$\beta\ra 0$ gives vanishing mutual information: this limit has
$\beta E_i\ra 0$ which is mathematically equivalent to the
$\theta_{1,2}=0$ subcase earlier, with $I[A,B]\ra 0$. In the
present context, this is $t\ra 0$, and we again
obtain vanishing mutual information, $I[A,B]\ra 0$.

\subsection{2-qutrit systems}

Consider now two qutrits, $|i\ran,\ i=0,1,2$: the Hamiltonian
(in eigenstate basis) and the normalized time evolution operator are
\bea
\quad\ H = \sum E_{ij} |ij\ran\lan ij|\,,\ &&\ 
E_{ij} = \{E_{00}, E_{11}, E_{22}, E_{01}, E_{02}, E_{12}\}\ ,\\
\rho_t = {e^{-iE_{ij}t}\over\sum_{ij} e^{-iE_{ij}t}}\, |ij\ran\lan ij|
\!\!&=&\!\! {e^{-iE_{ij}t} \over
e^{-iE_{00}t}+e^{-iE_{11}t}+e^{-iE_{22}t}+2e^{-iE_{01}t}+2e^{-iE_{02}t}+2e^{-iE_{12}t}}\,
|ij\ran\lan ij| ,\nn
\eea
again with $E_{ij}=E_{ji}$.\
The reduced time evolution operator tracing over the second
qutrit is
\be
(\rho_t^A)_{ij} = (\rho_t)_{ijkl} \delta^{kl}\,;\qquad
\rho_t^A = {1\over\sum_{ij} e^{-iE_{ij}t}}\ \sum_{i=0,1,2}
\big(\sum_j e^{-iE_{ij}t}\big) |i\ran\lan i|\ .
\ee
In general this leads to complex-valued entropy as before, with multiple
distinct phases. Imposing exchange symmetry between the qutrits, \ie\
$|0\ran\leftrightarrow |1\ran \leftrightarrow |2\ran$, this reduces
to a single independent phase controlled by $-(E_{01}-E_{00})t$ which then
gives real entropy.

\subsection{Two uncoupled oscillators}

We consider two uncoupled harmonic oscillators: the Hamiltonian is
\be
H = \sum E_{n_1 n_2}\, |n_1, n_2\ran \lan n_1, n_2|\,,\qquad
E_{n_1 n_2} = \omega (n_1+n_2+1)\ .
\ee
The normalized time evolution operator then becomes
\be\label{rhot-2osc}
\rho_t = \sum {e^{-iE_{n_1 n_2}t}\over \sum e^{-iE_{n_1 n_2}t}}\,
|n_1, n_2\ran\lan n_1, n_2|
\ee
The normalization evaluates to
\be\label{HO-normalization}
\sum_{1, 2} e^{-iE_{n_1 n_2}t}
= e^{-i\omega t} \sum_{1, 2} e^{-i\omega n_1t}\ e^{-i\omega n_2 t}
= {e^{-i\omega t} \over (1 - e^{-i\omega t})^2}\ .
\ee
Now, tracing over the second oscillator, we obtain
\be\label{rhotA-2osc}
\rho_t^A = \sum_{n_2=0}^\infty \rho_t
= \sum_{n_1} {e^{-i\omega n_1t}\over 1/(1-e^{-i\omega t})}\, |n_1\ran \lan n_1|
\ee
with the von Neumann entropy
\be\label{StA-2osc}
S_t^A = -\sum_n {e^{-i\omega nt}\over 1/(1-e^{-i\omega t})}\,
\log {e^{-i\omega nt}\over 1/(1-e^{-i\omega t})}
= -\log (1-e^{-i\omega t}) + {i\omega t\,e^{-i\omega t}\over 1-e^{-i\omega t}}\ ,
\ee
which is the usual entropy for a single oscillator at finite
temperature with $\beta=it$. In general this is complex-valued. The
zero temperature limit gives $S\sim \beta E\,e^{-\beta E}$ which here
gives $S\sim it \omega\,e^{-i\omega t}$\,.

In evaluating the normalization (\ref{HO-normalization}), it is
important to note that this sum over the infinite tower of states (and
similar quantities involving any infinite tower of states) is not
strictly convergent as an infinite series since this complex
expression is highly oscillatory for high energy states, although the
sum and its closed form expression are formally true. This is also
true for the single oscillator expression (\ref{rhotA-2osc}) obtained
as the reduced time evolution operator, whose normalization
is\ $\sum_{n_1}e^{-i\omega n_1t}=1/(1-e^{-i\omega t})$. Towards
rendering this well-defined as a series, one can introduce a small
regulator either in $\omega$ or in $t$ (giving time a tiny regulating
Euclidean component) which then makes it converge: \eg\ a small
Euclidean time component gives
\be\label{tE-HO-reg}
\sum_{n_1}e^{-i\omega n_1(t-i\epsilon)} =
\sum_{n_1}e^{-i\omega n_1t} e^{-n_1\omega\epsilon} 
= {1\over 1-e^{-i\omega (t-i\epsilon)}}\,,
\ee
which defines the sum. An alternative way
to view it is to start with the (convergent) finite temperature
partition function $\sum_ne^{-\beta E_n}$ and then perform analytic
continuation to imaginary temperature $\beta=it$.

It is interesting to also study two coupled harmonic oscillators with
Hamiltonian
\be
H=\frac{1}{2} \,  (p_A^2 + p_B^2 ) + \frac{k_1}{2} \, (x_A^2+ x_B^2)
\,+\, \frac{k_2}{2} \, (x_A- x_B)^2\ .
\ee
We describe this in detail in Appendix~\ref{app:cpldOsc}. The resulting
entropy from the time evolution operator can be realized as following
from imaginary temperature.

\section{The time evolution operator with projections}\label{sec:tE-proj}

As we have seen, the entanglement structures arising from the time
evolution operator involve the entire space of states since the time
evolution operator is like a full density matrix. It is desirable to
isolate a ``part'' of the time evolution operator, to understand
various components of the latter. This suggests appending projections
onto individual states.

With this in mind, we now consider the time evolution operator along
with a projection operator onto some state $|i\ran$, as in
(\ref{tE-pr}):
\be\label{tE-pr-2}
\rho_t^{|i\ran}
= {\rho_t|i\ran\lan i|\over {\rm Tr} (\rho_t|i\ran\lan i|)}
= {|f[i]\ran \lan i|\over {\rm Tr}(|f(i)\ran \lan i|)}\ ,\qquad
|f[i]\ran = e^{-iHt}|i\ran\ .
\ee
(The projection here is from the right: at the calculational level,
projecting from the left is similar but leads to complex conjugate
expressions in general.)
The state $|f[i]\ran$ is the final state obtained by time-evolving
the initial state $|i\ran$.\ If $|i\ran$ is a Hamiltonian eigenstate,
then $\rho_t^{|i\ran}$ reduces to just a single component $|i\ran\lan i|$
(the phase coefficient cancels upon normalizing), \ie\ the usual
density matrix for $|i\ran$. This is also true at $t=0$ for a generic
state $|i\ran$: here\
$\rho_t^{|i\ran}|_{t=0}={|i\ran\lan i|\over {\rm Tr} (|i\ran\lan i|)}$\
which gives ordinary entanglement structures at $t=0$.

For a generic state $|i\ran$, we obtain (\ref{tE2-tEp}).
As a simple concrete example, consider the 2-state system (\ref{qm2state})
earlier with a generic initial state:
\bea
|i\ran = c_1|1\ran+c_2|2\ran\ \ \ (|c_1|^2+|c_2|^2=1) \quad\ \ \ra\quad\ \
|f[i]\ran = c_1e^{-iE_1t}|1\ran+c_2e^{-iE_2t}|2\ran\,; \nn\\ [1mm]
\rho_t^{|i\ran} = {\cal N}^{-1} \Big(|c_1|^2e^{-iE_1t}|1\ran\lan 1| +
|c_2|^2e^{-iE_2t}|2\ran\lan 2| + c_1c_2^*e^{-iE_1t}|1\ran\lan 2|
+ c_2c_1^*e^{-iE_2t}|2\ran\lan 1| \Big),
\eea
where ${\cal N}={\rm Tr}(|f\ran\lan i|)$ is the normalization.
Now taking $|1\ran\equiv |++\ran$ and $|2\ran\equiv |--\ran$ and
performing a partial trace over the second component gives
\bea\label{rhotA|i>-2state}
&& \rho_t^{|i\ran, A} = {1\over |c_1|^2 + |c_2|^2e^{i\theta}}
\Big( |c_1|^2 |+\ran\lan +| + |c_2|^2e^{i\theta} |-\ran\lan -| \Big) ,\qquad
\theta=-(E_2-E_1)t\,, \nn\\   
&& S_t^{|i\ran, A} = -{|c_1|^2\over |c_1|^2 + |c_2|^2e^{i\theta}}
\log {|c_1|^2\over |c_1|^2 + |c_2|^2e^{i\theta}}
- {|c_2|^2e^{i\theta}\over |c_1|^2 + |c_2|^2e^{i\theta}}
\log {|c_2|^2e^{i\theta}\over |c_1|^2 + |c_2|^2e^{i\theta}}\,.\qquad 
\eea
At $t=0$, the von Neumann entropy above is ordinary entanglement
entropy for the generic state $|i\ran$ (obtained from $\rho_A={\rm
  Tr}_B\,|i\ran\lan i|$).  For general timelike separation $t$, the
entropy $S_A$ is real-valued only if $|c_1|^2=|c_2|^2$, \ie\ maximal
entanglement at $t=0$ (or $\theta=0$).

Consider now two qubits, each $|1\ran, |2\ran$, with a general
Hamiltonian (\ref{H-2qubit}) as before.
For a generic state
\be
|I\ran = \sum_{ij}c_{ij}|ij\ran\ ,
\ee
with the basis $|ij\ran=\{ |11\ran, |22\ran, |12\ran, |21\ran \}$, 
and the time evolution operator with projection can be evaluated as
(\ref{tE2-tEp}).
Performing a partial trace over the second component here gives
\bea\label{tE2-tEp-2}
\rho_t^{|I\ran, A} &=& {1\over \sum_{ij}|c_{ij}|^2e^{-iE_{ij}t}}
\sum_{i,k=1}^2 \big(\sum_j c_{ij}c_{kj}^*e^{-iE_{ij}t}\big)|i\ran\lan k| \nn\\
&=& {1\over \sum_{ij}|c_{ij}|^2e^{-iE_{ij}t}} \Big(
\big(|c_{11}|^2e^{-iE_{11}t}+|c_{12}|^2e^{-iE_{12}t}\big) |1\ran\lan 1|  \nn\\
&& \qquad\ +\
\big(c_{11}c_{21}^*e^{-iE_{11}t}+c_{12}c_{22}^*e^{-iE_{12}t}\big) |1\ran\lan 2| +\
\big(c_{21}c_{11}^*e^{-iE_{12}t}+c_{22}c_{12}^*e^{-iE_{22}t}\big) |2\ran\lan 1| \nn\\
&& \qquad\qquad\qquad\quad +\
\big(|c_{21}|^2e^{-iE_{12}t}+|c_{22}|^2e^{-iE_{22}t}\big) |2\ran\lan 2| \Big)
\eea
At $t=0$, this is ordinary entanglement for the generic state $|I\ran$.
There are special subcases with interesting structure, some of which
we will discuss soon.

For 3-qubits with Hamiltonian (\ref{H-3qubitchain}) with energies
$E_{ijk}$ for eigenstates $|ijk\ran$ (alongwith the symmetry-based
simplifications there), we obtain
\bea
|I\ran=\sum_{i,j,k=1}^2c_{ijk}|ijk\ran : &&
\rho_t^{|I\ran} = {1\over \sum_{ijk}|c_{ijk}|^2e^{-iE_{ijk}t}}
\sum_{i,j,k,l,m,n=1}^2 c_{ijk}c_{lmn}^*e^{-iE_{ijk}t}|ijk\ran\lan lmn|\,,\nn\\
\rho_t^{|I\ran, A} &=& {1\over \sum_{ijk}|c_{ijk}|^2e^{-iE_{ijk}t}}
\sum_{j,m=1}^2 \big(\sum_i\sum_k c_{ijk}c_{imk}^*e^{-iE_{ijk}t}\big)|j\ran\lan m|\,,
\eea
where the last line is the reduced transition matrix for the middle
qubit, arising after a partial trace over the 1st and 3rd components
$(\rho_t^A)_{jm}=(\rho_t)_{ijk,lmn}\delta^{il}\delta^{kn}$.

\subsection{Thermofield-double type states}\label{sec:tE-proj-TFD}

It is interesting to focus on thermofield-double type initial states
with only ``diagonal'' components: then for 2-qubits, using
(\ref{tE2-tEp-2}) we obtain
\bea\label{tE-proj-TFD}
&& |I\ran = \sum_{i=1,2} c_{ii}|ii\ran\,:\qquad\
\rho_t^{|I\ran} = {1\over \sum_i|c_{ii}|^2e^{-iE_{ii}t}}
\sum_{i,k=1}^2 c_{ii}c_{kk}^*e^{-iE_{ii}t}|ii\ran\lan kk| ,\nn\\
&& \rho_t^{|I\ran, A} 
= {1\over |c_{11}|^2e^{-iE_{11}t}+|c_{22}|^2e^{-iE_{22}t}}
\Big( |c_{11}|^2e^{-iE_{11}t}|1\ran\lan 1|
+ |c_{22}|^2e^{-iE_{22}t}|2\ran\lan 2| \Big)\,.
\eea
This is identical to (\ref{rhotA|i>-2state}). 
To elaborate a little, the initial state is\
$|I\ran=c_{11}|11\ran+c_{22}|22\ran$ and its time-evolved final state is\
$|F\ran=c_{11}e^{-iE_{11}t}|11\ran+c_{22}e^{-iE_{22}t}|22\ran$, and
the reduced time evolution operator with projection, $\rho_t^{|I\ran, A}$
above, is the normalized reduced transition matrix for $|I\ran,\ |F\ran$,
with the corresponding (in general complex-valued) pseudo-entropy
(\ref{tEpE}).

Now restricting further to maximally entangled states with
$|c_{11}|^2=|c_{22}|^2={1\over 2}$ simplifies this to just a single
nontrivial phase $e^{i\theta}=e^{-i\Delta E\,t}$ where $\Delta E=E_{22}-E_{11}$,
thereby leading to the entanglement structure (\ref{rhotA-2state})
of the time evolution operator for the 2-state case, \ie\
$S_t^{|I\ran, A} = -{1\over 1+e^{i\theta}} \log {1\over 1+e^{i\theta}}
- {1\over 1+e^{-i\theta}} \log {1\over 1+e^{-i\theta}}$. 
The states in question here can be regarded as maximally entangled
Bell pairs and the entropy can be regarded as pseudo-entropy
for the Bell pair initial state $|I\ran$ and its time-evolved final
state $|F\ran$.  As noted there, this is a real-valued entropy,
oscillating in time with periodicity set by $\Delta E$, growing
unbounded at specific time values where $t={(2n+1)\pi\over \Delta E}$.
Note also that specific time values $t={2n\pi\over\Delta E}$ lead to
the minimum value $S_A=\log 2$, which is simply the ordinary
entanglement entropy of the maximally entangled initial state. The
fact that this time entanglement entropy can be unbounded is a novel
feature compared with ordinary entanglement entropy for ordinary
quantum systems.

For an $n$-qubit system comprising basis states
$|\{i_1,\ldots,i_n\}\ran$, with $i_k=1,2$, the time evolution operator
with projection onto generic initial states gives complicated
entanglement structure. However projecting onto thermofield
double type initial states, we obtain
\be
|I\ran=\sum_{i=1,2} c_{ii\ldots i} |ii\ldots i\ran\,:\qquad
\rho_t^{|I\ran, A} = {1\over \sum_{i}|c_{ii\ldots i}|^2e^{-iE_{ii\ldots i}t}}
\sum_{i=1}^2 |c_{ii\ldots i}|^2e^{-iE_{ii\ldots i}t} |i\ran\lan i|\,,
\ee
which is identical to the 2-qubit case. It is clear that any qubit
system has identical entanglement structure for the time evolution
operator with projection onto thermofield double type states.
Now if we additionally restrict to maximal entanglement, we have
both $|c_{ii\ldots i}|^2$ equal so $|c_{ii\ldots i}|^2={1\over 2}$\,.
This again contains just one nontrivial phase thereby leading to
the entanglement structure of the time evolution operator for the
2-state case, \ie\ (\ref{rhotA-2state}).

\section{Time evolution operator, normalized at $t=0$}\label{sec:tE-normt=0}

In this section, we will discuss aspects of the time evolution
operator with normalization at $t=0$ (rather than at general time $t$),
following \cite{Narayan:2022afv}. This gives
\be\label{rhot(t)1}
\rho_t^0(t) \equiv {{\cal U}(t)\over {\rm Tr}\,{\cal U}(0)}
\quad\ra\quad \rho_t^{0,A} = tr_B\,\rho_t\quad\ra\quad
S_A = -tr (\rho_t^A\log\rho_t^A)\ .
\ee
The normalization ensures that we obtain ordinary entanglement structures
at $t=0$.\ In this case ${\rm Tr}\,\rho_t(t)=1$ at $t=0$ but not at
general $t$. This gives quite different entanglement structures, as
we will see.

Since ${\cal U}(0)=\sum_I|I\ran\lan I|={\bf 1}$ \ie\ the identity operator
made up as a sum over all eigenstate projection operators, the
normalization factor is ${\rm Tr}\,{\cal U}(0)=N$, the dimension of
the Hilbert space, constant in time. Thus for a general bipartite
system we obtain
\be
\rho_t^0(t) = {1\over N} \sum_{i,i'} e^{-iE_{i,i'}t} |i,i'\ran\lan i,i'|\quad
\ra\quad \rho_t^{0,A}
= {1\over N}\sum_i \big(\sum_{i'} e^{-iE_{i,i'}t}\big) |i\ran\lan i|\,,
\ee
differing from (\ref{tE1-tE}) only in the normalization.\ 
A general 2-qubit system (\ref{H-2qubit}) now gives
\be
\rho_t^0(t) = {1\over 4} \sum_{ij} e^{-iE_{ij}t}\,|ij\ran\lan ij|
\ee
and taking a partial trace over the second component gives
\bea\label{rhot0A-2qubit}
&& \rho_t^{0,A} = {1\over 4} \Big( \big(e^{-iE_{11}t}+e^{-iE_{12}t}\big)
|1\ran\lan 1| + \big(e^{-iE_{21}t}+e^{-iE_{22}t}\big) |2\ran\lan 2| \Big) \nn\\
&& S_t^{0,A} = -{1\over 4}\big(e^{-iE_{11}t}+e^{-iE_{12}t}\big)
\log\Big({1\over 4}\big(e^{-iE_{11}t}+e^{-iE_{12}t}\big)\Big)\, \nn\\
&& \qquad\qquad\qquad\ -\, {1\over 4}\big(e^{-iE_{21}t}+e^{-iE_{22}t}\big)
\log\Big({1\over 4}\big(e^{-iE_{21}t}+e^{-iE_{22}t}\big)\Big) .
\eea
In general $S_t^{0,A}$ is a complicated complex entropy. However there
are special cases. If all energy values are the same, this simplifies to
\bea
&& E_{ij}=E_0:\qquad \rho_t = {e^{-iE_{0}t}\over 4} \sum_{ij} |ij\ran\lan ij|\,,
\quad \rho_t^{0,A} = {e^{-iE_{0}t}\over 2} \sum_{i=1,2} |i\ran\lan i|\,, \nn\\
&&\quad S_t^{0,A} = -e^{-iE_0t} \log \Big({1\over 2} e^{-iE_0t}\Big)
= \left(\log 2 + iE_0t\right)\,e^{-iE_0t}\ .
\eea

Appending a projection operator for a state $|i\ran$ as in
sec.~\ref{sec:tE-proj}, we obtain
\be\label{pE-norm0}
\rho_t^{0,|i\ran} = {\rho_t^0\,|i\ran\lan i|\over
  {\rm Tr}(\rho_t^0\,|i\ran\lan i|)}
= {|f[i](t)\ran\lan i|\over {\rm Tr}\,({\cal U}(0)|i\ran\lan i|)}
= {|f[i](t)\ran\lan i|\over {\rm Tr} (|i\ran\lan i|)}\,,
\ee
since ${\cal U}(0)$ is the identity operator.
This is similar to  (\ref{tE-pr}), but differs in normalization.
So if the initial state is unit-normalized, the normalization factor
is a trivial $1$. This is not ordinary entanglement even if the state
is an eigenstate since the nontrivial time evolution phase remains.
For instance a 2-qubit system (\ref{H-2qubit}) gives
\be
|i\ran=|11\ran\,:\quad \rho_t^{0,|i\ran}
= {{\cal U}(t)|11\ran\lan 11|\over {\rm Tr} (|11\ran\lan 11|)}
= e^{-iE_{11}t}|11\ran\lan 11|\,,
\ee
after projecting onto a simple eigenstate $|11\ran$.\
The partial trace then gives
\be\label{pE-norm0-eigen}
\rho_t^{0,|i\ran,A} = {\rm Tr}_{_2}\rho_t^{0,|i\ran} = e^{-iE_{11}t}|1\ran\lan 1|
\ \ \Rightarrow\ \ S_t^{0,|i\ran,A} = -e^{-iE_{11}t}\,\log\big(e^{-iE_{11}t}\big)
= iE_{11}t\,e^{-iE_{11}t}\,.
\ee
The normalization at $t=0$ makes this different from ordinary mixed
state entanglement structures at finite temperature, although these
still resemble imaginary temperature structures. Although it might
seem natural to normalize at general $t$, part of the motivation here,
following \cite{Narayan:2022afv}, is that the time evolution only
enters via the final state in (\ref{pE-norm0}), which apart from this
is akin to the pseudo-entropy (\ref{tE-pr}), (\ref{tE-pr-2}). This
appears to help isolate the timelike characteristics, as in
(\ref{pE-norm0-eigen}) where the leading time-dependence is manifestly
pure imaginary: it would be interesting to explore this further.

\section{2-dim CFTs and timelike intervals}\label{sec:2dCFT-tE}

The studies of $dS_3$ extremal surfaces in \cite{Doi:2022iyj,Narayan:2022afv}
and \cite{Hikida:2022ltr,Hikida:2021ese}, led to studies of timelike
entanglement in ordinary 2-dim CFT (in particular (\ref{tE-2dCFT})):
we now elaborate on this\ (there are parallels with some discussions
in \cite{Doi:2023zaf} which appeared as we were finalizing this paper).

We want to consider the time evolution operator as a density operator
towards exploring entanglement-like structures: towards this we define
\be
\rho_t[\{\phi(x)\}|\{\phi(x')'\}]
= {1\over Z_t}\,\lan\{\phi(x)\}|e^{-itH}|\{\phi(x')\}\ran
\ee
with $Z_t={\rm Tr}\ e^{-itH}$. However rendering this well-defined
is best done in the Euclidean path integral formulation, defining
the ground state wavefunction for the configuration $\phi(x')$ as
\be
\Psi[\{\phi(x')\}] = \int_{t_E=-\infty}^{\phi(t_E=0,x)=\phi(x')} D\phi\, e^{-S_E}
= \int_{t_E=-\infty}^{t_E=0} D\phi\, e^{-S_E}\,
\prod_x \delta(\phi(t_E=0,x)-\phi(x'))
\ee
with $S_E$ the Euclidean action for the field $\phi(t_E,x)$\
(we model this discussion along the lines of 
\cite{Holzhey:1994we,Calabrese:2004eu,Calabrese:2009qy}, and
\cite{Ryu:2006ef}). Now the
reduced density matrix for the interval $A$ is obtained from
$\rho_t[\phi_0(x)|\phi_0'(x')]$ above by performing a partial trace
over the environment $B$ setting $\phi_0(x)=\phi_0'(x)$. This becomes
\be\label{rho-xx'}
\rho[\phi(x)_{0^+}|\phi(x)_{0^-}] = {1\over Z}
\int_{t_E=-\infty}^{t_E=\infty} D\phi\,e^{-S_E(\phi)}\,
\prod_{x\in A}\delta(\phi(0^+,x)-\phi(x)_{0^+})\
\delta(\phi(0^-,x)-\phi(x)_{0^-})
\ee
In this form there is no sacrosanct meaning to what we define as
Euclidean time: the differences for a timelike interval only enter in
the analytic continuation to Lorentzian signature eventually. For
a free massless 2-dim scalar, the action is\
$S_E=\int dt_E dx\,((\del_{t_E}\phi)^2 + (\del_x\phi)^2)$
and Euclidean evolution appears symmetric between $t_E, x$.
For the usual spacelike interval, the reduced density matrix involves
Euclidean time evolution along $t_E$: for a timelike interval on the
other hand, the reduced density matrix involves Euclidean time
evolution along $x$ which is regarded as Euclidean time now
calculationally. So we have
\be\label{rhot-tEtE'}
\rho_t[\phi(t_E)_{0^+}|\phi(t_E)_{0^-}] = {1\over Z_{t_E}}
\int_{x=-\infty}^{x=\infty} D\phi\,e^{-S_E(\phi)}\,
\prod_{t_E\in A}\delta(\phi(t_E,0^+)-\phi(t_E)_{0^+})\
\delta(\phi(t_E,0^-)-\phi(t_E)_{0^-})
\ee
Apart from $x\leftrightarrow t_E$, this is equivalent to (\ref{rho-xx'}).

Let us now discuss this in terms of Hamiltonians for a free massless
scalar: note that Euclidean and Lorentzian times are related as
$t_E=it$. For the usual time coordinate $t$, the Hamiltonian is\
$H_t^+=\int dx\,((\del_t\phi)^2 + (\del_x\phi)^2)=
\int dx\,(-(\del_{t_E}\phi)^2 + (\del_x\phi)^2)$: this is positive
definite. Now compactifying $t_E$ can be used to obtain the reduced
density matrix ${\rm Tr}_B\,e^{-\beta_tH}$ at finite temperature for
an interval with width $\Delta x$.
With $x$ taken as Euclidean time, we obtain the Hamiltonian
$H_x=\int dt_E\,((\del_{t_E}\phi)^2 - (\del_x\phi)^2)$. Now
compactifying $x$ with periodicity $\beta_x$ and considering a timelike
interval with width $\Delta t$, the reduced density matrix becomes
\bea
&& H_x = \int dt_E\, (-(\del_x\phi)^2+(\del_{t_E}\phi)^2)
= -i\int dt\, ((\del_x\phi)^2+(\del_{t}\phi)^2) \equiv -iH_x^+\,;\nn\\
&&\qquad\qquad\qquad\qquad\qquad
\rho_t^A = {\rm Tr}_B\,e^{-\beta_x H_x} = {\rm Tr}_B\,e^{i\beta_x H_x^+}\,,
\eea
so that in terms of the positive definite Hamiltonian $H_x^+$, this
resembles a thermal reduced density matrix but with imaginary temperature.

The usual replica formulation of entanglement entropy for a single
interval proceeds by picking some Euclidean time direction $\tau_E$
and the interval $\Delta x\equiv [u,v]$ on that slice, then
constructing $n$ replica copies of the space glued at the interval
endpoints and evaluating ${\rm Tr}\rho_A^n$. The reduced density
matrix for the ground state is formulated as above, via Euclidean time
evolution, with appropriate boundary conditions for the fields on the
replica sheets. Then ${\rm Tr}\rho_A^n$ in the replica space can be
mapped to the twist operator 2-point function at the interval endpoints
which implement the boundary conditions across the sheets. This
finally leads to
\be
S_A = -\lim_{n\ra 1} \del_n{\rm Tr}\rho_A^n\ \ra\
{c\over 6}\log {(\Delta x)^2\over \epsilon^2}\ .
\ee
The only data that enters this is the central charge of the CFT and
the interval in question. When we consider a timelike interval, the
above formulation goes through with the only change being that the
Euclidean time slice we pick is the spatial slice $x=const$ with
the interval being $\Delta t\equiv [u_t,v_t]$. However now when we
continue back to Lorentzian time, we must rotate $u_t, v_t$ accordingly,
so the spacetime interval is
\be
\Delta^2 = -(\Delta t)^2 = -(v_t-u_t)^2\ ,
\ee
and the entanglement entropy becomes
\be\label{tE-2dCFT}
S_A = {c\over 6}\log {\Delta^2\over \epsilon^2}
= {c\over 6}\log {-(\Delta t)^2\over \epsilon^2}
= {c\over 3}\log {\Delta t\over \epsilon} + {c\over 6}(i\pi)\ ,
\ee
with the imaginary part arising as $i\pi=\log(-1)$.
Note that imaginary values also arise in studies of quantum extremal
surfaces in de Sitter with regard to the future boundary
\cite{Chen:2020tes,Goswami:2021ksw}, stemming from
timelike-separations.

The discussions above are formulated in terms of Euclidean path
integrals with an eventual analytic continuation to obtain timelike
interval entanglement. Along the lines of our finite quantum system
descriptions, one could consider Lorentzian time evolution explicitly.
Towards this consider a CFT on a cylinder, with time running along
the axis. The Hamiltonian is\
$H_{cyl} = {\pi\over l} (L_0+{\bar L}_0-{c+{\bar c}\over 24})$\ and
the unnormalized time evolution operator becomes $e^{-iH_{cyl}t}\sim
q^{\sum_n nN_n} |N_n\ran\lan N_n|$ with $q=e^{-2it/l}$ for both left/right
modes, and the normalization becomes
${\rm Tr}\, q^{\sum_n nN_n} = \prod_{n=1}^\infty {1\over 1-q^n}$
(the ${c+{\bar c}\over 24}$ factor cancels with normalization).
In the momentum basis, the time evolution operator is an infinite
sum of decoupled oscillators. Recalling the case of two uncoupled
oscillators (\ref{rhot-2osc}), tracing out all
higher mode oscillators leaving only the lowest frequency $n=1$
oscillator mode naively gives\
$\rho_t^A = \sum_{n} {q^n\over 1/(1-q)}\, |n\ran \lan n|$ and
$S_t^A = -\log (1-q) - {q\log q\over 1-q}$, with appropriate limits
as described after (\ref{rhotA-2osc}).
Also, along the lines of sec.~\ref{sec:tE-proj}, we can study
aspects of the time evolution operator along with projection onto
initial states. We leave these and related investigations for the future.

\section{Time entanglement, time-dependent interactions}\label{sec:t-depInt}

So far we have considered time-independent Hamiltonians. In these
cases we can relate the time evolution operator to the thermal density
matrix by the analytic continuation $\beta \ra it$, consistent with
the expectation that time independence maps to thermal equilibrium. In
this section, we consider some special simple examples of
time-dependent Hamiltonians: we expect that the time evolution
operator will not admit any simple map to some thermal density matrix
in such cases (no thermal equilibrium).

We obtain the time evolution operator in the interaction picture by
solving the Schrodinger time evolution equations, evolving the state
by the time evolution operator 
\begin{equation}\label{state_evolution_time_evoulution_interaction_subspace}
  \ket{\alpha,t\,;t_0}_I=U_I(t,t_0)\,\ket{\alpha,t_0;t_0}_I
  =\sum\, c_{ij}(t)\ket{ij} \,.
\end{equation} 
This enables to determine the time evolution operator, where $\ket{ij}$
are the eigenstates of $H_0$ (and $t_0=0$). Our conventions are those
of \cite{Sakurai}, with the interaction picture time evolution equations
of the form\ $i\hbar{d\over dt} c_N(t)=\sum_M V_{NM} e^{i\omega_{NM}t} c_M(t)$
with $\omega_{NM}=E_N-E_M$.

As a toy example, consider a 2-state system with states $|1\ran, |2\ran$,
and energies $E_1, E_2$: then a $\delta$-function interaction
$V_{12}=V\delta(t-\epsilon)$ (with $\epsilon>0$ an infinitesimal regulator)
gives the interaction picture evolution equations\
(with ${\dot c}_i={d\over dt}c_i$) 
\bea\label{2state-deltafn}
i\hbar{\dot c_1}=V_{12}e^{i\omega_{12}t}c_2\,,\ &&\
i\hbar{\dot c_2}=V_{21}e^{i\omega_{21}t}c_1\,;\nn\\
i\hbar c_1(t)=Vc_2(\epsilon)+i\hbar c_1(0)\,,\ &&\
i\hbar c_2(t)=Vc_1(\epsilon)+i\hbar c_2(0)\,,
\eea
where the second line is obtained by integrating across the interaction
support at $t=\epsilon$ (and the phases $e^{i\omega_{12}t}$ are trivial). Since
the time dependence is
only nontrivial for $t=\epsilon$, we see that $c_i(t)=c_i(\epsilon)$, \ie\ the
coefficients remain unchanged for $t\geq \epsilon$.
Solving for $c_1(t),c_2(t)$ gives\
$\big(^{c_1(t)}_{c_2(t)}\big)=\rho_{t,I} \big(^{c_1(0)}_{c_2(0)}\big)$\
with generic initial state $c_1(0),c_2(0)$,
where the interaction picture time evolution operator is\
$\rho_{t,I}={1\over 1+{V^2\over\hbar^2}} \big(|1\ran\lan 1|
+ {V\over i\hbar}|1\ran\lan 2| + + {V\over i\hbar}|2\ran\lan 1| +
|2\ran\lan 2|\big)$\ 
(this can also be seen to agree with time dependent perturbation theory).
We now generalize this sort of delta-function coupling interaction
to a system of two qubits to study time entanglement.

Consider a simple system of two qubits with the time-dependent interaction
\begin{equation}\label{time_dependent_interaction_subspace}
  V_I(t)= V \delta{(t-\epsilon)}\,
  \big(\ket{11}\bra{12}\,+\,\ket{12} \bra{11}\big)\,,
\end{equation}
with an infinitesimal regulator $\epsilon>0$ (so the impulse
interaction is just after $t=0$). The Hamiltonian $H_0$ before turning
on the interaction ($t\leq0$) has eigenstates
$\ket{11}$, $\ket{22}$, $\ket{12}$, $\ket{21}$, and eigenvalues
$E_{11}$, $E_{22}$, $E_{12}$, $E_{21}=E_{12}$, respectively.
The time evolution equations for the coefficients (suppressing the
phases), and their integrated versions, are (with $\hbar=1$)
\bea\label{differential_equations_for _coefficients_C's_subspace}
\dfrac{d}{dt} c_{11}(t)=-iV\,\delta{(t-\epsilon)}\, c_{12}(t)\,\,\,,\ &&\
\dfrac{d}{dt} c_{12}(t)=-iV\,\delta{(t-\epsilon)}\,c_{11}(t)\,, \nonumber \\
\dfrac{d}{dt} c_{21}(t)=0 \,\qquad\qquad,\ &&\ \dfrac{d}{dt} c_{22}(t)=0\,,
\nn\\ [3mm]
\!\!\! \Rightarrow\qquad\quad
c_{11}(t)=c_{11}(0)-\,iV\,c_{12}(\epsilon) \quad ,\ &&\
c_{12}(t)=c_{12}(0)-\,iV\,c_{11}(\epsilon),\nonumber \\ 
c_{21}(t)=c_{21}(0)\, \qquad\qquad,\ &&\ c_{22}(t)=c_{22}(0)\,.
\eea
We now note that the $c_{ij}(t)=c_{ij}(\epsilon)$ for the impulse
interaction, where $t\geq \epsilon$, since there is no nontrivial time
dependence after $t=\epsilon$. This then gives
\begin{align}\label{coefficients_C(t)'s_subspace} 
& c_{11}(t)=\frac{1}{1+V^2}\,\Big(c_{11}(0)-\, iV\,c_{12}(0)\Big)\,\,\,,\qquad c_{12}(t)=\frac{1}{1+V^2}\,\Big(c_{12}(0)-\, iV\,c_{11}(0)\Big), \nonumber \\
& c_{21}(t)=c_{21}(0)\,\,\hspace{4.4cm},\qquad c_{22}(t)=c_{22}(0)\,.
\end{align} 
This gives the interaction picture time evolution operator $U_I(t,t_0)$
(with $t_0=0$ and $t>0$) which maps\
$\big(^{c_{11}(t)}_{c_{12}(t)}\big)= U_I(t) \big(^{c_{11}(0)}_{c_{12}(0)}\big)$
in the $\{|11\ran, |12\ran\}$ subspace,\
using (\ref{state_evolution_time_evoulution_interaction_subspace}).
Then the time evolution operator $U(t)\equiv {\tilde\rho}_t$ in the
Schr$\ddot{o}$dinger picture is\ (with $\rho_t$ the normalized one)
\begin{align}\label{time_evol_operator_schrodinger_picture_subspace_diracdelta_potential}
{\tilde\rho}_t=e^{-iH_0t}\, U_I(t) 
&=\frac{1}{1+V^2}\,\Big( e^{-iE_{11}t}\,\ket{11}\bra{11}\,-\,iV e^{-iE_{11}t}\,\ket{11}\bra{12}\,-\,iV\, e^{-iE_{12}t}\,\ket{12}\bra{11}\, \nonumber \\
&\hspace{1.5cm}+\,e^{-iE_{12}t}\,\ket{12}\bra{12}\,\Big)\,+\,e^{-iE_{12}t}\,\ket{21}\bra{21}\,+\, e^{-iE_{22}t}\,\ket{22}\bra{22}\,,\nn\\
\rho_t={\cal N}_V{\tilde\rho}_t\,,\qquad
& \quad {\cal N}_V^{-1} \equiv
{\rm Tr}({\tilde\rho}_t) = \frac{1}{1+V^2} \big(e^{-iE_{11}t}+e^{-iE_{12}t}\big)
+ e^{-iE_{12}t}+e^{-iE_{22}t} \,.
\end{align}
We now find the reduced time evolution operator by tracing 
out a qubit.\ $\rho_t^{A}$ arises from tracing out the second qubit in
$\rho_t$, and $\rho_t^{B}$ from tracing out the first qubit:
\begin{align}
& {\cal N}_V^{-1} \rho_t^{A}=\frac{1}{1+V^2}\,\Big( e^{-iE_{11}t}\,+\,e^{-iE_{12}t}\Big) \ket{1}\bra{1}\,+\,\Big( e^{-iE_{12}t}\,+\,e^{-iE_{22}t}\,\Big)\ket{2}\bra{2}\,,
  \nonumber  \\
& {\cal N}_V^{-1} \rho_t^{B}=\frac{1}{1+V^2}\,\Big( e^{-iE_{11}t}\,\ket{1}\bra{1}\,-\,iV e^{-iE_{11}t}\,\ket{1}\bra{2}\,-\,iV\, e^{-iE_{12}t}\,\ket{2}\bra{1}\, \nonumber \\ 
& \hspace{3.5cm}+\,e^{-iE_{12}t}\,\ket{2}\bra{2}\,\Big)\,+\,e^{-iE_{12}t}\,\ket{1}\bra{1}\,+\,e^{-iE_{22}t}\,\ket{2}\bra{2}\, \label{reduced_density_matrix_tracing_first_qubit__interacting_hamiltonian_subspace}.    
\end{align}
Note that $\rho_t^{A}=\rho_t^{B}$ for $V=0$ is in agreement with
sec.~\ref{sec:tE-examples} for the 2-qubit system.
The entropy associated with $\rho_t^{A}$ or $\rho_t^{B}$ is
complex-valued in general.

Consider now the same 2-qubit system but a more general impulse interaction
\begin{equation}\label{V_I(t)-moreGenImpulse}
V_I(t)= V \delta{(t-\epsilon)}\,\big( \ket{11}\bra{12}+\ket{12} \bra{11}+\ket{21}\bra{22}+\ket{22}\bra{21} \big)\,.
\end{equation}
Using (\ref{state_evolution_time_evoulution_interaction_subspace}),
the interaction picture time evolution equations and the integrated
versions are
\bea\label{differential_equations_for_coefficients_C's_dirac_delta_interaction}
\dfrac{d}{dt} c_{11}(t)=-iV\,\delta{(t-\epsilon)}\,c_{12}(t) \,\,,\ &&\
\dfrac{d}{dt} c_{12}(t)=-iV\,\delta{(t-\epsilon)}\,c_{11}(t) \,, \nonumber \\
\dfrac{d}{dt} c_{21}(t)=-iV\,\delta{(t-\epsilon)}\,c_{22}(t)\,\,,\ &&\
\dfrac{d}{dt} c_{22}(t)=-iV\,\delta{(t-\epsilon)}\,c_{21}(t)\,,
\nn\\ [3mm]
\!\!\! \Rightarrow\qquad\quad
c_{11}(t)=c_{11}(0)-\,iV\,c_{12}(\epsilon)\,\,,\ &&\
c_{12}(t)=c_{12}(0)-\,iV\,c_{11}(\epsilon) \,, \nonumber \\
c_{21}(t)=c_{21}(0)-\,iV\,c_{22}(\epsilon)\,\,,\ &&
c_{22}(t)=C_{22}(0)-\,iV\,c_{21}(\epsilon) \,.
\eea
These are the analogs for the interaction (\ref{V_I(t)-moreGenImpulse}) 
of (\ref{differential_equations_for _coefficients_C's_subspace}) with
the simpler interaction (\ref{time_dependent_interaction_subspace}).
As before, we have $c_{ij}(t)=c_{ij}(\epsilon),\ t\geq \epsilon$, 
since there is no nontrivial time dependence after the impulse at
$t=\epsilon$. Solving for $c_{ij}(t)$ leads 
here to the Schr$\ddot{o}$dinger picture time evolution operator
$U(t)\equiv{\tilde\rho}_t$  (with $\rho_t$ the normalized one)
\begin{align}\label{time_evol_operator_time_interaction_schr_picture_dirac_delta_interaction}
{\tilde\rho}_t=e^{-iH_0t}\, U_I(t)
&=\frac{1}{1+V^2}\,\Big( e^{-iE_{11}t}\,\ket{11}\bra{11}\,-\,iV e^{-iE_{11}t}\,\ket{11}\bra{12}\,-\,iV\, e^{-iE_{12}t}\,\ket{12}\bra{11} \nonumber \\
& \hspace{2.4cm}+\,e^{-iE_{12}t}\,\ket{12}\bra{12}\, +\,e^{-iE_{12}t}\,\ket{21}\bra{21}\,-\,iV e^{-iE_{12}t}\,\ket{21}\bra{22}\, \nonumber \\
& \hspace{2.4cm}-\,iV\, e^{-iE_{22}t\,}\,\ket{22}\bra{21}\,+\,e^{-iE_{22}t}\,\ket{22}\bra{22}\,\Big)\,,\nn\\
\rho_t={\cal N}_V{\tilde\rho}_t\,\qquad
& \quad {\cal N}_V^{-1} \equiv
{\rm Tr}({\tilde\rho}_t) = \frac{1}{1+V^2} \big(e^{-iE_{11}t}+2e^{-iE_{12}t}
+e^{-iE_{22}t}\big) \,.
\end{align}
Tracing out either the second qubit or the first gives $\rho_t^{A}$
or $\rho_t^{B}$:
\begin{align}
\rho_t^{A} &={\cal N}_V\,\frac{1}{1+V^2}\,\Big(\,( e^{-iE_{11}t}\,+\,e^{-iE_{12}t}) \ket{1}\bra{1}\,+\, (e^{-iE_{12}t}\,+\,e^{-iE_{22}t})\,\ket{2}\bra{2}\Big)\,, \nonumber   \\  
\rho_t^{B} &={\cal N}_V\,\frac{1}{1+V^2}\,\Big( (e^{-iE_{11}t}\,+\,e^{-iE_{12}t})\,\ket{1}\bra{1}\,-\,iV (e^{-iE_{11}t\,}\,+\,e^{-iE_{12}t} )  \ket{1}\bra{2}\, \nonumber \\
& \hspace{2.5cm} -\,iV (e^{-iE_{12}t\,}\,+\,e^{-iE_{22}t} ) \ket{2}\bra{1}\,+\,(e^{-iE_{12}t}\,+\,e^{-iE_{22}t})\,\ket{2}\bra{2}  \Big)\,. \label{reduced density matrix_tracing_first_qubit_ interaction_dirac_delta}
\end{align}
Note that here the ${1\over 1+V^2}$ factors cancel with that in ${\cal N}_V$
(which is an accident; this would not occur if the interaction strengths
in (\ref{V_I(t)-moreGenImpulse}) were not uniformly $V$ for all matrix
elements).\
As for (\ref{reduced_density_matrix_tracing_first_qubit__interacting_hamiltonian_subspace}), we see that these reduced time evolution
operators are equal, $\rho_t^{A}=\rho_t^{B}$, for $V=0$, in agreement with
sec.~\ref{sec:tE-examples}. These give complex-valued entropy in general,
although there are special cases with real entropy:
\eg\ for $E_{11}=E_{22}=E_{12}$ we obtain
$\rho_t^B={1\over 2}(^{\ \ 1}_{-iV} {}^{-iV}_{\ \ 1 })$ with eigenvalues
$\lambda_k={1\over 2}(1\pm iV)$\,: then the entropy
$S_t^B=-\sum_k \lambda_k\log\lambda_k$ becomes real-valued giving\
$S_t^B=\log 2 -{1\over 2}(1+iV)\log(1+iV)-{1\over 2}(1-iV)\log(1-iV)$.

We now look at this time evolution operator with projection onto some
initial state, along the lines of sec.~\ref{sec:tE-proj}. First consider
a thermofield-double type initial state $|I\ran=\sum_{i=1,2} c_{ii}|ii\ran$
as in sec.~\ref{sec:tE-proj-TFD}:\ this gives (with ${\cal N}$ the
normalization)
\bea
{\cal N} \rho_t|I\ran\lan I| = {{\cal N}\over 1+V^2}
    \big(\rho_t|I\ran\lan I|\big)\big|_{V=0}
- {\cal N}\,{iV\,e^{-iE_{12}t}\over 1+V^2} \Big( |c_{11}|^2 |12\ran\lan 11|
+ c_{11}c_{22}^* |12\ran\lan 22| \nn\\
+\ c_{11}^*c_{22} |21\ran\lan 11|
+ |c_{22}|^2 |21\ran\lan 22| \Big) .
\eea
A partial trace over the second or first qubit gives, respectively,
\bea
&& \rho_{t,I}^A = {\cal N}\,{1\over 1+V^2}\, \rho_t^A\big|_{V=0}
- {\cal N}\,{iV\,e^{-iE_{12}t}\over 1+V^2} \Big( c_{11}c_{22}^* |1\ran\lan 2|
+ c_{11}^*c_{22} |2\ran\lan 1| \Big)\,,\nn\\
&& \rho_{t,I}^B = {\cal N}\,{1\over 1+V^2}\, \rho_t^B\big|_{V=0}
- {\cal N}\,{iV\,e^{-iE_{12}t}\over 1+V^2} \Big( |c_{11}|^2 |2\ran\lan 1|
+ |c_{22}|^2 |1\ran\lan 2| \Big)\,.
\eea
This thus leads to nontrivial contributions to the complex-valued
entropy stemming from the impulse interaction controlled by the strength
$V$. For special cases the entropy is real: \eg\
$E_{11}=E_{22}=E_{12}$ with maximally entangled initial state
$c_{11}=c_{22}={1\over\sqrt{2}}$ gives
$\rho_{t,I}^A=\rho_{t,I}^B={1\over 2}(^{\ \ 1}_{-iV} {}^{-iV}_{\ \ 1 })$
with eigenvalues $\lambda_k={1\over 2}(1\pm iV)$ leading to real
entropy\ $S_t^B=-\sum_k \lambda_k\log\lambda_k$.

This is essentially the pseudo-entropy for the initial state
$|I\ran=c_{11}|11\ran+c_{22}|22\ran$ and its time evolved final state
using ${\tilde\rho}_t$ in (\ref{time_evol_operator_time_interaction_schr_picture_dirac_delta_interaction})\
\be
|F\ran={\tilde\rho}_t|I\ran
=\frac{1}{1+V^2}\Big(e^{-iE_{11}t}c_{11}|11\ran+e^{-iE_{22}t}c_{22}|22\ran
-iV\,e^{-iE_{12}t} c_{11}|12\ran -iV\,e^{-iE_{12}t} c_{22}|21\ran\Big)\,.
\ee

If on the other hand, one considers some initial state within the
$\{|11\ran, |12\ran \}$ subspace, then it turns out that
$\rho_{t,I}^A\propto |1\ran\lan 1|$ while $\rho_{t,I}^B$ has
eigenvalues $0, 1$\ (perhaps this is not surprising since any state in
this subspace is of a factorized form $|1\ran_A(a|1\ran+b|2\ran)_B$).
This leads to vanishing pseudo entropy for $\rho_{t,I}^A$ and
$\rho_{t,I}^B$\,.

We have illustrated the time evolution operator and its time entanglement
structure focussing on simple 2-qubit examples involving an impulse
$\delta$-function interaction. We have obtained the time evolution
operator by solving the time evolution Schrodinger equation for the
state coefficients. The time-dependence of the interaction leads to
nontrivial dependence on the interaction strength $V$, in addition
to the dependence on the energy eigenvalues and the timelike
separation $t$. No simple continuation via some imaginary temperature
exists here, unlike the discussions in the rest of the paper with
time-independent quantum systems. It is likely that general
time-dependent quantum systems will exhibit similar features. Perhaps
there are deeper ways to formulate timelike entanglement, which
make more explicit a partial trace over time paths or histories.

\section{Discussion}

We have studied various aspects of entanglement like structures with
timelike separations arising from the time evolution operator regarded
as a density operator, following \cite{Narayan:2022afv}. There are
close parallels with pseudo-entropy \cite{Nakata:2020luh} as we have
seen. The entropy from the time evolution operator alongwith
projection onto some initial state as we have seen in
sec.~\ref{sec:tE-proj} is identical to pseudo-entropy for the initial
state and its time-evolved final state. More broadly, there are large
parallels of the investigations here and in \cite{Narayan:2022afv}
with corresponding ones in \cite{Doi:2022iyj,Doi:2023zaf}. In general
the non-Hermitian structures here give complex-valued entropy,
although there are several interesting real-valued subfamilies
\eg\ (\ref{rhotA-2state}),\ special subcases of (\ref{rhotA-2qubit})
and (\ref{tE-proj-TFD}),\ qubit chains App.~\ref{app:qubitChains} with
the $|1\ran\leftrightarrow|2\ran$ exchange symmetry, and so on. The
behaviour of this entropy is quite different from usual spatial
entanglement entropy: for instance, (\ref{rhotA-2state}) oscillates in
time and appears to grow large at specific time
values. Correspondingly at other specific periodic time values the
entropy acquires its minimum value, coinciding with ordinary
entanglement entropy for the initial state (see
sec.~\ref{sec:tE-proj-TFD} in the context of thermofield-double
states, akin to Bell pair states). Overall these appear to be new
entanglement-like measures involving timelike separations, likely with
many new aspects open for exploring further. (It is also worth noting
other work
\eg\ \cite{Jonay:2018yei,Castellani:2021zrk,Lerose:2021svg,Diaz:2021snw},
which may have bearing on this broad circle of ideas.)

While more detailed understanding and physical interpretation of time
entanglement in general is yet to be developed, the mapping to
pseudo-entropy allows certain connections to previously studied
quantities. Pseudo-entropy stems from the transition matrix ${\cal
  T}_{F|I}$ in (\ref{tEpE}), (\ref{tE3-pE}), regarded as a generalized
density operator involving a preparation state and a postselected
state. Related quantities pertain to weak values of operators,
obtained as ${\cal O}_w={\rm Tr} ({\cal T}_{F|I} {\cal O})$. These are
in general complex-valued, not surprising since the transition matrix
is not a hermitian object (unlike ordinary hermitian density
matrices). See \eg\ \cite{dresselWeak,SalekPostselection} for more on
postselected states, conditional entropy and weak values (including
some experimental aspects). In the current context, components of the
time evolution operator can be isolated via projections onto specific
initial states as we have seen in sec.~\ref{sec:tE-proj}: this then
maps onto the corresponding pseudo-entropy. Thus time entanglement
with projection onto initial state $|I\ran$ dovetails with
postselected states being the corresponding time-evolved states. We
hope to obtain more refined understanding of these interrelations in
the future.

The finite quantum systems we have studied allow analysis using
Hamiltonian eigenstates and are thus intrinsically straightforward.
Time-independent Hamiltonians allow mapping the time evolution
operator to a thermal density matrix by the analytic continuation
$\beta \ra it$, consistent with the expectation that time independence
can be mapped to thermal equilibrium.  We expect that in cases with
nontrivial time dependence, these time-entanglement structures will
become more intricate with no natural imaginary temperature analytic
continuation: along the lines of studies of scattering amplitudes, we
expect that analogs of the interaction picture will be useful in
organizing these time entanglement structures. All these are
vindicated in the simple 2-qubit examples with $\delta$-function
impulse potentials (sec.~\ref{sec:t-depInt}), where we solve
explicitly for the nontrivial time evolution operator and the
corresponding time entanglement structures.  Related, complementary
studies (including holographic ones) appear in \cite{Nakata:2020luh},
\cite{Doi:2022iyj}, \cite{Mollabashi:2020yie}-\cite{Chen:2023gnh}.  We
hope to report further on these in the future.

We now make a few remarks on de Sitter extremal surfaces anchored at
the future boundary, which have timelike components, in particular
paraphrasing some discussions in \cite{Narayan:2023zen}. The $dS/CFT$
dictionary \cite{Maldacena:2002vr} $Z_{CFT}=\Psi_{dS}$ implies that
boundary entanglement entropy is bulk pseudo-entropy (since a replica
formulation on $Z_{CFT}$ amounts to one on $\Psi_{dS}$, \ie\ single
ket rather than a density matrix). Among other things this leads to
novel entropy relation/inequalities based on the complex-valued $dS$
extremal surface areas. This is put in perspective by comparing with
time-entanglement/pseudo-entropy in qubit systems, using the analyses
in this paper, in particular sec.~\ref{sec:tE-proj}: this reveals
striking differences for mutual time-information, tripartite
information and strong subadditivity (see sec.2.5 in
\cite{Narayan:2023zen}). The $dS$ areas give definite signs for these
quantities relative to those obtained from
time-entanglement/pseudo-entropy for qubit systems (with the final
state being time-evolved from the initial state). Since the $dS$ areas
are analytic continuations from $AdS$, these differences are perhaps
not surprising in light of the studies in \cite{Hayden:2011ag} (which
reveal definite signs the $AdS$ RT surface area inequalities compared
with those for entanglement entropy in qubit systems), but they are
striking.  Overall there are new entanglement structures here stemming
from timelike separations: we expect that the investigations here and
related ongoing ones will lead to further insights into both quantum
information and holography.

\vspace{8mm}

{\footnotesize \noindent {\bf Acknowledgements:}\ \ It is a pleasure
  to thank Ronak Soni and Tadashi Takayanagi for helpful discussions
  and comments on a draft.  This work is partially supported by a
  grant to CMI from the Infosys Foundation.  }


\appendix

\section{Time evolution, pseudo-entropy: special cases}\label{app:tE-pE}

Consider now the pseudo-entropy transition matrix (\ref{tEpE}) for the
2-state case (\ref{qm2state}), with arbitrary initial state $|i\ran$
and arbitrary final state $|f\ran$,
\bea
&& |i\ran=c_1|1\ran+c_2|2\ran\,,\quad |f\ran=c_1'|1\ran+c_2'|2\ran\,; \nn\\
&& {\cal T}_{f|i} = {1\over c_1'c_1^*+c_2'c_2^*}
\Big(c_1'c_1^*|1\ran\lan 1| + c_2'c_2^*|2\ran\lan 2|
+ c_1'c_2^*|1\ran\lan 2| + c_2'c_1^*|2\ran\lan 1| \Big)\,.\ \ 
\eea
With $|1\ran\equiv |++\ran, |2\ran\equiv |--\ran$, a partial trace
over the second component gives 
\be
{\cal T}_{f|i}^A = {1\over c_1'c_1^*+c_2'c_2^*}
\Big(c_1'c_1^*|+\ran\lan +| + c_2'c_2^*|-\ran\lan -| \Big)
\ee
as the reduced transition matrix.
To compare with entanglement for the time evolution operator, we take
the final state to be time-evolved from some other initial state
$|i'\ran$ so
\be\label{TF|I^A-2state}
|f\ran=c_1'e^{-iE_1t}|1\ran+c_2'e^{-iE_2t}|2\ran\quad\ra\quad
{\cal T}_{f|i}^A = {\big(c_1'c_1^*|+\ran\lan +|
  + c_2'c_2^*e^{i\theta}|-\ran\lan -| \big)
  \over c_1'c_1^*+c_2'c_2^*e^{i\theta}}\,,
\ee
with $\theta=-(E_2-E_1)t$. Then we see that:\\
$\bullet$\ \ using (\ref{rhotA-2state}) for the time
evolution operator, ${\cal T}_{f|i}^A=\rho_t^A$\ if\
$c_1=c_1'={1\over\sqrt{2}}\,,\ c_2=c_2'={1\over\sqrt{2}}$,\ \ie\
the initial and final states are identical maximally entangled states.\\
$\bullet$\ \ using (\ref{rhotA|i>-2state}) for the time evolution
operator with projection,\ ${\cal T}_{f|i}^A=\rho_t^{|i\ran}$\ if\
$c_1'=c_1 ,\ c_2'=c_2$,\ \ie\ $|f\ran=|f[i]\ran$ \ie\ the final state
is time-evolved from the initial state $|i'\ran=|i\ran$.

This structure of mapping ${\cal T}_{f|i}^A=\rho_t^A$ however is not
true more generally. For instance, consider two qubits more generally,
as in (\ref{H-2qubit}).
Then the pseudo-entropy transition matrix (\ref{tEpE}) becomes
\be
|I\ran=\sum_{i,j=1}^2c_{ij}|ij\ran\,,\quad |F\ran=\sum_{i,j=1}^2c_{ij}'|ij\ran\,;
\qquad
{\cal T}_{F|I} = {1\over \sum_{ij} c_{ij}'c_{ij}^*}
\sum_{i,j,k,l=1}^2 c_{ij}'c_{kl}^*\,|ij\ran\lan kl|
\ee
and partial trace over the 2nd component gives the reduced transition
matrix as
\bea\label{TF|I^A-2qubit}
&& {\cal T}_{F|I}^A = {1\over \sum_{ij} c_{ij}'c_{ij}^*}
\sum_{i,k=1}^2 (\sum_jc_{ij}'c_{kj}^*)\,|i\ran\lan k| 
\ =\ {1\over \sum_{ij} c_{ij}'c_{ij}^*} \Big(
(c_{11}'c_{11}^*+c_{12}'c_{12}^*)|1\ran\lan 1| + \nn\\
&& \qquad\qquad (c_{11}'c_{21}^*+c_{12}'c_{22}^*)|1\ran\lan 2| 
 +\ (c_{21}'c_{11}^*+c_{22}'c_{12}^*)|2\ran\lan 1| +
(c_{21}'c_{21}^*+c_{22}'c_{22}^*)|2\ran\lan 2| \Big) .\qquad
\eea
Towards comparing with the time evolution operator, we think of the
future state as time-evolved from some initial state, \ie\ 
$|F\ran = \sum_{ij} c_{ij}'e^{-iE_{ij}t}|ij\ran$.
It is then clear that pseudo-entropy via the reduced transition
matrix matches time entanglement via the normalized time evolution
operator with projection onto $|i\ran$, \ie\
${\cal T}_{f|i'}^A = \rho_t^{|i\ran,A}$\ if the final state is taken
to be time-evolved from the initial state, \ie\
$|F\ran={\cal U}(t)|I\ran$ so $c_{ij}'=c_{ij}e^{-iE_{ij}t}$.
However, in contrast with (\ref{TF|I^A-2state}), the fact that there
are off-diagonal terms in (\ref{TF|I^A-2qubit}) makes the structure
different from the reduced time evolution operator. 
To set the off-diagonal terms to vanish, we could consider specializing
to maximally entangled thermofield-double type initial and final states,
and with $|F\ran$ time-evolved from $|I\ran$,\ \ie\
$|I\ran = \sum_{ii} c_{ii} |ii\ran$ with
$c_{ij}, c_{ij}'=0,\ i\neq j,\ c_{ii}=c_{jj}\ \forall\ i,j$,\
and\ $|F\ran = \sum_{ii} c_{ii}' |ii\ran = {\cal U}(t)|I\ran$.\
In this case, we find that all the off-diagonal terms vanish and we
obtain the reduced transition matrix to be of the same form as in
(\ref{TF|I^A-2state}). On the other hand the reduced time evolution
operator for the general 2-qubit case is (\ref{rhotA-2qubit}), which
has two distinct phases in general. Thus the reduced transition
matrix differs from the reduced time evolution operator. One can
engineer special energy values $E_{ij}$ where the two coincide
(although this appears ad hoc).

Of course, these structures are with a single Hilbert space for
constructing both initial and final states. Doubling the Hilbert
spaces directly enables a map from the transition matrix to the time
evolution operator in general, as in sec.~\ref{sec:tEp-TM}.

\section{Qubit chains}\label{app:qubitChains}

Now we consider qubit chains to understand time entanglement
structures.
For any nearest neighbour 2-qubit pair, we impose nearest-neighbour
interactions, with
\bea\label{2qubit-Ha1a2}
&& \qquad\qquad\qquad 
s|q\ran = a_q|q\ran\,,\quad |q\ran=\{ |1\ran, |2\ran\}\,;\qquad
H = -Js_1s_2\,,\nn\\ [1mm]
&& H[11]=E_{11}=-Ja_1^2\,,\quad H[22]=E_{22}=-Ja_2^2\,,
\quad H[12]=H[21]=E_{12}=-Ja_1a_2\,.\qquad
\eea
In the first line, we are defining operators $s_i$ with action as above
(the $i$ being the site label), that give the qubit Hamiltonian action
elaborated on in the second line.
This Hamiltonian generalizes the 2-qubit case (\ref{H-2qubit}) earlier.
(Imposing a $|1\ran\leftrightarrow|2\ran$ exchange symmetry simplifies
this to Ising-like interactions, as we will discuss later.)

{\bf 3-qubit chain:}\ \ Consider now a chain of 3 qubits with
Hamiltonian based on the nearest neighbour 2-qubit interaction above.
This gives the 3-qubit chain Hamiltonian as
\bea\label{H-3qubitchain}
&& \qquad\qquad\qquad\  H = -J(s_1s_2+s_2s_3)\ ,\nn\\
&& H \equiv E_I|I\ran\lan I| = E_1|111\ran\lan 111| + E_2|222\ran\lan 222|
+ E_5 \big( |121\ran\lan 121| + |212\ran\lan 212| \big) \nn\\
&& \qquad\qquad\qquad\qquad\
+\ E_3 \big( |112\ran\lan 112| + |211\ran\lan 211| \big)
+ E_4 \big( |122\ran\lan 122| + |221\ran\lan 221| \big)\,,\qquad \nn\\
&& E_1=-2Ja_1^2=2E_{11}\,,\quad E_2=-2Ja_2^2=2E_{22}\,,\quad
E_5=-2Ja_1a_2=2E_{12}\,, \nn\\
&& E_3=-Ja_1^2-Ja_1a_2=E_{11}+E_{12}\,,\quad
E_4=-Ja_1a_2-Ja_2^2=E_{22}+E_{12}\,,\\
&& \qquad
E_4-E_3={1\over 2} (E_2-E_1)\,,\quad E_1+E_5=2E_3\,,\quad E_2+E_5=2E_4\,.\nn
\eea
Then the time evolution operator ${\cal U}(t)$ after normalizing
becomes
\be
\rho_t = {1\over e^{-iE_1t}+e^{-iE_2t}+2e^{-iE_3t}+2e^{-iE_4t}+2e^{-iE_5t}}\,
\sum_I e^{-iE_It}|I\ran\lan I| \equiv
{\cal N}\,\sum_I e^{-iE_It}|I\ran\lan I|\ .
\ee
Now tracing out the 1st and 3rd qubit states gives the reduced time
evolution operator 
\be
(\rho_t^A)_{11} = {\cal N}\,
\big( e^{-iE_1t} + 2 e^{-iE_3t} + e^{-iE_5t} \big)\ ,\qquad
(\rho_t^A)_{22} = {\cal N}\,
\big( e^{-iE_2t} + 2 e^{-iE_4t} + e^{-iE_5t} \big)\ ,
\ee
for the middle qubit. Using the relations between the $E_i$ in
(\ref{H-3qubitchain}) simplifies this to
\bea\label{rhotA-3qubit}
&& (\rho_t^A)_{11} = {\cal N}\,\big( e^{-iE_{11}t} + e^{-iE_{12}t} \big)^2\,,\qquad
(\rho_t^A)_{22} = {\cal N}\,\big( e^{-iE_{22}t} + e^{-iE_{12}t} \big)^2\,,\nn\\
&& \qquad\qquad\ {\cal N}^{-1} = {\rm Tr}\,{\cal U}(t)
=  \big( e^{-iE_{11}t} + e^{-iE_{12}t} \big)^2 +
\big( e^{-iE_{22}t} + e^{-iE_{12}t} \big)^2\ .
\eea
In general, this is a function of three independent parameters
$E_{11}, E_{22}, E_{12}$\ (or equivalently $E_1, E_2, E_5$) so it is a
complex-valued function of three phases in general. A straightforward
real slice is obtained when there is a
$|1\ran\leftrightarrow|2\ran$ exchange symmetry as we will discuss
later.

{\bf 5-qubit chain:}\ \ the configurations and their energies are
\bea\label{5qubit-configs}
&& |11111\ran,\ \ 4E_{11};\qquad |22222\ran,\ \ 4E_{22};\qquad
|12121\ran,\ \ |21212\ran,\ \ 4E_{12}; \nn\\
&& |11112\ran,\ \ |11122\ran,\ \ |11222\ran,\ \ |12222\ran,\ \ 3E_{11}+E_{12};
\nn\\
&& |22221\ran,\ \ |22211\ran,\ \ |22111\ran,\ \ |21111\ran,\ \ 3E_{22}+E_{12};
\nn\\
&& |11121\ran,\ \ |11211\ran,\ \ |12111\ran,\ \ |21112\ran,\ \
2E_{11}+2E_{12}; \nn\\
&& |12221\ran,\ \ |22212\ran,\ \ |22122\ran,\ \ |21222\ran,\ \
2E_{22}+2E_{12}; \nn\\
&& |11221\ran,\ \ |12211\ran,\ \ |22112\ran,\ \ |21122\ran,\ \
E_{11}+E_{22}+2E_{12};\quad \nn\\
&& |11212\ran,\ \ |12112\ran,\ \ |21211\ran,\ \ |21121\ran,\ \
E_{11}+3E_{12};\nn\\
&& |12122\ran,\ \ |12212\ran,\ \ |22121\ran,\ \ |21221\ran,\ \
E_{22}+3E_{12};\quad 
\eea
Tracing over all but the middle (3rd) qubit gives the reduced
time evolution operator as
\bea\label{rhotA-5qubit}
&& ({\tilde\rho}_t)^{(3)}_{11}
= e^{-i(4E_{11})t}+e^{-i(4E_{12})t}+2e^{-i(3E_{11}+E_{12})t}+
2e^{-i(3E_{22}+E_{12})t}+2e^{-i(E_{11}+E_{22}+2E_{12})t} \qquad \nn\\
&& \qquad\qquad +\,3e^{-i(2E_{11}+2E_{12})t}+e^{-i(2E_{22}+2E_{12})t}
+2e^{-i(E_{11}+3E_{12})t}+2e^{-i(E_{22}+3E_{12})t}\,, \nn\\
&& ({\tilde\rho}_t)^{(3)}_{22}
= e^{-i(4E_{22})t}+e^{-i(4E_{12})t}+2e^{-i(3E_{22}+E_{12})t}+
2e^{-i(3E_{11}+E_{12})t}+2e^{-i(E_{11}+E_{22}+2E_{12})t} \nn\\
&& \qquad\qquad +\,3e^{-i(2E_{22}+2E_{12})t}+e^{-i(2E_{11}+2E_{12})t}
+2e^{-i(E_{22}+3E_{12})t}+2e^{-i(E_{11}+3E_{12})t}\,,
\eea
where the tilde denotes un-normalized.
The normalization of the time evolution operator here becomes
\be
{\cal N}_5^{-1} = {\rm Tr}\,{\tilde\rho}_t^{(3)} = {\rm Tr}\,{\cal U}(t)
   = ({\tilde\rho}_t)^{(3)}_{11} + ({\tilde\rho}_t)^{(3)}_{22}
\ee
In general the resulting von Neumann entropy is a complicated
complex-valued function of the three energy parameters
$E_{11}, E_{22}, E_{12}$.

There are parallels between our discussions here on qubit chain
configurations and those in \cite{Jatkar:2017jwz} on ghost-spin
chains (although the context is different).

\bigskip

{\bf Infinite qubit chain:}\ \ Consider now an infinite 1-dim chain
of qubits, again with only nearest-neighbour interactions, the
Hamiltonian being
\be
H = -J\sum_n s_n s_{n+1} = \ldots - Js_{-1}s_0 - Js_0s_1 + \ldots
\ee
We can focus on the qubit at location $n=n_0$ as the subsystem
in question, tracing over all the other qubits in the chain. 
The reduced time evolution operator is
\be
\rho_t = {1\over \sum_I e^{-iE[I]t}} \sum_{n_0=1,2}
\big(\sum_{I;\ n\neq 0} e^{-iE[I]t}\big) |n_0\ran\lan n_0|
\ee
This is a complicated object in general, although still simply a
complex-valued function of the three energy parameters $E_{11}, E_{22}, E_{12}$.
Since this qubit only interacts directly with its two neighbours,
the effective system has some parallels with the 3-qubit chain above:
but the detailed structure is complicated, as already evident in
the 5-qubit case earlier.

\medskip

{\bf $|1\ran\leftrightarrow|2\ran$ exchange symmetry:}\ \
In the simple subcase enjoying $|1\ran\leftrightarrow|2\ran$ exchange
symmetry, there are substantial simplifications in (\ref{2qubit-Ha1a2}):
this is when there is an Ising-like structure, with
\be\label{1-2symmE}
a_1=-a_2=1\,;\qquad  E_{11}=E_{22}=-E_{12}=-J\ .
\ee
For instance the 3-qubit case (\ref{rhotA-3qubit}) simplifies to
\be
{\cal N}_3^{-1} =  2\big( e^{iJt} + e^{-iJt} \big)^2 \,, \qquad\
(\rho_t^A)_{11} = (\rho_t^A)_{22} =
{\cal N}_3\,\big( e^{iJt} + e^{-iJt} \big)^2 = {1\over 2}\ ,
\ee
which thus gives von Neumann entropy $\log 2$. Likewise the 5-qubit
(\ref{rhotA-5qubit}) case can be seen to simplify to
\be
{\cal N}_5^{-1} =  2\big( e^{iJt} + e^{-iJt} \big)^4 \,, \qquad\
(\rho_t^A)_{11} = (\rho_t^A)_{22} =
{\cal N}_5\,\big( e^{iJt} + e^{-iJt} \big)^4 = {1\over 2}\ ,
\ee
so the middle qubit has identical structure. For an infinite qubit
chain with this Ising-like $\BZ_2$ symmetry, we expect translation
invariance in the ``bulk'' so we expect that the reduced time evolution
operator has again similar structure. Considering an $N$-qubit chain
(towards large $N$), the configurations can be organized similar to
(\ref{5qubit-configs}).  It is then clear that the ground states are
$|11\ldots 11\ran,\ |22\ldots 22\ran$, with energy $-(N-1)J$. The first
excited states comprise ``one kink'' states with exactly one $12$- or
$21$-interface with energy $-(N-3)J$ and degeneracy $2(N-1)$. The next
set of excited states contain two kinks, so the energy is
$-(N-5)J$ with degeneracy $4(N-2)$.\ Higher excited states
contain multiple $12$- or $21$-interfaces. The two highest energy
states have maximally alternating $1,2$s, \ie\ $|12121..\ran,
|21212..\ran$: there are $(N-1)$ interfaces giving energy $(N-1)J$.
Furthermore, every energy $E$ (with corresponding configurations)
comes in pairs, \ie\ there are corresponding configurations with
energy $-E$. This can be seen above, with the ground states and
highest energy states: likewise, corresponding to the one kink states,
we have states with energy $(N-3)J$ obtained by transforming one
of the $12$- or $21$-interfaces in the highest energy states to
$11$ or $22$, which then lowers the energy precisely by $2J$ (and
their degeneracy can be checked easily). Thus the normalization
of the time evolution operator (akin to the partition function) is
${\cal N}_N^{-1} = {\rm Tr}\,{\tilde\rho}_t$, \ie\
\be
{\cal N}_N^{-1} 
= 2\big(e^{iJt(N-1)}+(N-1)e^{iJt(N-3)}+\ldots+(N-1)e^{-iJt(N-3)}+e^{-iJt(N-1)}\big)
  = 2\big( e^{iJt} + e^{-iJt} \big)^{N-1}\ .
\ee
Each component of the reduced time evolution operator for some bulk
qubit can be explicitly seen to receive contributions equally from
half these states: so we obtain
\be
(\rho_t^A)_{11} = (\rho_t^A)_{22} =
{\cal N}_N\,\big( e^{iJt} + e^{-iJt} \big)^{N-1} = {1\over 2}\ ,
\ee
which is identical to the structure of the middle qubit in the
previous finite qubit cases.

Note that it is adequate to require $E_{11}=E_{22}$ to implement this
$|1\ran\leftrightarrow|2\ran$ exchange symmetry: then shifting the
energies arrives at the symmetric values in (\ref{1-2symmE}).  However
if keep $E_{12}$ independent of $E_{11}=E_{22}$ then there are
apparently two independent parameters: however it is straightforward
to see that the reduced time evolution operator, wbile non-Hermitian,
nevertheless leads to real-valued von Neumann entropy. It is likely
that similar studies can be extended for ``ghost-spin'' models
such as those in \cite{Narayan:2016xwq,Jatkar:2017jwz}.

All of the above structures can be seen to match ordinary finite
temperature entanglement, except with imaginary temperature
$\beta=it$.

\section{Two coupled oscillators}\label{app:cpldOsc}

We consider the following Hamiltonian $H$ with unit masses $m_A=m_B=1$ ,   
\begin{equation} \label{hamil_coupled_os}
H=\frac{1}{2} \,  (p_A^2 + p_B^2 ) + \frac{k_1}{2} \, (x_A^2+ x_B^2) \,+\, \frac{k_2}{2} \, (x_A- x_B)^2 \,\,.  
\end{equation}
This is slightly different from the coupled oscillators case discussed
in \cite{Nakata:2020luh}.
We diagonalise the Hamiltonian in a coordinate basis $\{y_1,y_2\}$ as
below. Then the hamiltonian (\ref{hamil_coupled_os}) becomes   
\bea \label{hamil_two_cos_ycoords}
&& H=(\,\frac{1}{2} \,p_1^2 + \frac{1}{2}\, \Omega_1^2 \, y_1^2 \,) + (\,\frac{ 1}{2} \, p_2^2\,+\,  \frac{1}{2}\, \Omega_2^2 \, y_2^2 \,) \,,   \nn\\
&&\quad y_1=\frac{(x_A+x_B)}{\sqrt{2}} \,\, ; \quad\
y_2=\frac{(x_A-x_B)}{\sqrt{2}}  \ ,
\eea
where $\Omega_1= \sqrt{k_1}$\,,\, $\Omega_2= \sqrt{k_1+2 k_2}$. 
The energy eigenvalues and eigenfunctions of (\ref{hamil_two_cos_ycoords}) are labelled by $E_{n_1n_2}$, and $\phi_{n_1 n_2}(y_1,y_2) $ respectively, 
\begin{align} 
E_{n_1n_2}= (n_1+\frac{1}{2}) \Omega_1  + (n_2+\frac{1}{2}) \Omega_2 \,=\, E_{n_1}+E_{n_2}\,;\quad \phi_{n_1n_2}(y_1,y_2)&= \phi_{n_1}(y_1) \, \phi_{n_2}(y_2) \,, \label{eneigen_N _eigenvec}   
\end{align}
where $n_1$,$n_2$ take values from $0$ to $\infty$ and
$E_{n_1}= (n_1+\frac{1}{2}) \Omega_1$\,,
$E_{n_2}= (n_2+\frac{1}{2}) \Omega_2 \,$.

We now write the time evolution operator in its eigenbasis as follows 
\begin{equation} \label{time_evol_operator_cos}
 e^{-iHt}=\rho(t)=\sum_{n_1,n_2}  \, e^{-  i\, E_{n_1 n_2} t} \, \ket{\phi_{n_1 n_2}}   \bra{\phi_{n_1 n_2}}  \ .
\end{equation}
In position space  
\begin{align}
\rho(y_1,y_2;y'_1,y'_2,t)&= \sum_{n_1,n_2}  \, e^{-  i\, E_{n_1 n_2} t} \, \phi_{n_1 n_2}(y_1,y_2) \, {\phi^*_{n_1 n_2}}(y'_1,y'_2)  \,\,\,\,\,, \nonumber \\
 &= \sum_{n_1,n_2}  \, e^{-  i\, (E_{n_1} +E_{n_2}) t} \, \phi_{n_1 n_2}(y_1,y_2) \, {\phi^*_{n_1 n_2}}(y'_1,y'_2)  \,\,, \,\,\, \nonumber \\
&=\rho_1(y_1;y'_1,t) \, \rho_2(y_2;y'_2,t) \ . \label{time_evol_dm_operator_ycoords}
  \end{align} 
We have applied (\ref{eneigen_N _eigenvec})  in the first line of (\ref{time_evol_dm_operator_ycoords}), and
\begin{align}
\rho_1(y_1;y'_1,t)= \sum_{n_1} \, e^{-iE_{n_1}t} \, \phi_{n_1}(y_1) \phi^*_{n_1}(y'_1) \, ;\ \
\rho_2(y_2;y'_2,t)= \sum_{n_2} \, e^{-iE_{n_2}t} \, \phi_{n_2}(y_2) \phi^*_{n_2}(y'_2) \,\label{rho1_rho2_matrix_operators_y_coords} \, .
\end{align}
(\ref{time_evol_dm_operator_ycoords}) shows that the time evolution operator $\rho(t)$ is decomposed as $\rho(t)=\rho_1(t) \otimes \rho_2(t)$. 
The energy eigenstate for a single harmonic oscillator of frequency $\Omega$
(setting $m=1$) is
\begin{equation} \label{Single_Harmonic_os_energyeigen_eigenvec}
\phi_n(x)= \frac{1}{\sqrt{2^n  n!}} \, \left(\frac{\Omega}{\pi}\right)^{\frac{1}{4}} \, e^{- \frac{\Omega \, x^2}{2} } \, H_n(\sqrt{\Omega} \, x) \,;\qquad E_n=(n+\frac{1}{2}) \Omega \,.
\end{equation}
We now use Mehler's formula for Hermite polynomials \cite{merzbacher}
\begin{equation} \label{Hermite_poly_Melher_formula}
\sum_{n=0}^{\infty} \, \frac{(\frac{\alpha}{2})^n}{n!} H_n(X) H_n(Y) \, = \frac{1}{\sqrt{1-\alpha^2}}\, e^{\frac{-\alpha^2 \left(X^2+Y^2\right)+2 \alpha XY   }{1-\alpha^2}}      \,.
\end{equation}
We now consider the time evolution operator for a single harmonic oscillator of frequency $\Omega$ in order to calculate (\ref{time_evol_dm_operator_ycoords}):
\begin{equation} \label{rho_single_harmonic_os}
\rho(x;x',t)=\sum_{n=0}^{\infty} \, e^{-iE_{n}t} \, \phi_{n}(x) \, \phi^*_{n}(x') \,\,. 
\end{equation}
Applying (\ref{Single_Harmonic_os_energyeigen_eigenvec}) into (\ref{rho_single_harmonic_os})
\begin{equation} \label{rho_single_harmonic_os_simp}
\rho(x;x',t)= \sum_{n=0}^{\infty} \, e^{-i (n+\frac{1}{2}) \Omega t} \, \frac{1}{2^n \,  n!}\,  \left(\frac{\Omega}{\pi}\right)^{\frac{1}{2}} e^{- \frac{\Omega}{2}\, (x^2+x'^2)}\, H_n(\sqrt{\Omega\,} x)\, H_n(\sqrt{\Omega}\, x') \ .
\end{equation} 
We now use (\ref{Hermite_poly_Melher_formula}) in (\ref{rho_single_harmonic_os_simp}),
 \begin{equation} \label{rho_single_harmonic_os_simp_1}
\rho(x;x',t)=\frac{ \left(\frac{\Omega}{\pi}\right)^{\frac{1}{2}}} {\sqrt{2 i \, \sin(\Omega\, t)}}   \,\,  e^{- \frac{p\, (x^2+x'^2)}{2}\,+ q \, xx'}\,,
\end{equation}
where
\begin{equation} \label{Expressions_p(t)Nq(t)}
p(t)= -i \, \Omega \, \cot(\Omega\,t)\,\,;\qquad q(t)=  \frac{-i \, \Omega} {\, \sin(\Omega\,t)} \, .
\end{equation} 
We will not write the $t$ dependence of $p$ and $q$ explicitly, we simply write $p$ and $q$ instead of $p(t)$ and $q(t)$. We now define the normalised time evolution operator as $P(x;x',t)= \frac{\rho(x;x',t)}{{\rm Tr}(\rho(x;x',t))} \,,$  
\begin{equation} \label{Normalised_rho_single_harmonic_os_simp_1}
P(x;x',t)= \sqrt{\frac{p-q}{\pi}} \,   e^{- \frac{p\, (x^2+x'^2)}{2}\,+ q \, xx'} \,.
\end{equation}
Note that the normalization ${\rm Tr}(\rho(x;x',t))$ using
(\ref{rho_single_harmonic_os_simp_1}) is\
$\int_{-\infty}^\infty dx\, \rho(x,x,t)$, which is oscillatory (rather
than a damped Gaussian), using (\ref{Expressions_p(t)Nq(t)}). To render
this well-defined, we insert a small exponentially damping regulator:
this is the position space analog of the regularization in
(\ref{tE-HO-reg}). Similar regulators are required to define various
infinite sums/integrals here.

We now find the expressions for $\rho_1(y_1;y'_1,t)$  and $\rho_2(y_2;y'_2,t)$ appearing in (\ref{time_evol_dm_operator_ycoords}) using (\ref{rho_single_harmonic_os_simp_1}),
\begin{align}
\rho_1(y_1;y'_1,t)=\frac{ \left(\frac{\Omega_1}{\pi}\right)^{\frac{1}{2}}} {\sqrt{2 i \, \sin(\Omega_1\, t)}}   \,\,  e^{- \frac{p\, (y_1^2+{y'_1}^2)}{2}\,+ q \, y_1 y'_1}\,, \nonumber \\
\rho_2(y_2;y'_2,t)=\frac{ \left(\frac{\Omega_2}{\pi}\right)^{\frac{1}{2}}} {\sqrt{2 i \, \sin(\Omega_2\, t)}}   \,\,  e^{- \frac{r\, (y_2^2+{y'_2}^2)}{2}\,+ s \, y_2 y'_2}\,, 
\end{align}  
where
\begin{align}
  p= -i \, \Omega_1 \, \cot(\Omega_1\,t)\,; \ \ \
  q=  \frac{-i \, \Omega_1} {\, \sin(\Omega_1\,t)} \,; 
  \ \ \ r= -i \, \Omega_2 \, \cot(\Omega_2\,t)\,;\ \ \
  s=  \frac{-i \, \Omega_2} {\, \sin(\Omega_2\,t)} \,.
  \label{Expressions_p(t)_q(t)_r(t)_s(t)} 
\end{align} 
We define the  normalised time evolution operator as $P(y_1,y_2;y'_1,y'_2,t)= \frac{ \rho(y_1,y_2;y'_1,y'_2,t)}{Tr(\rho(y_1,y_2;y'_1,y'_2,t))}$\,, \,    
\begin{equation} \label{Normalised_time_evol_dm_operator_ycoords}
P(y_1,y_2;y'_1,y'_2,t)= \sqrt{\frac{p-q}{\pi}} \,  \sqrt{\frac{r-s}{\pi}}\,    e^{- \frac{p\, (y_1^2+{y'_1}^2)}{2}\,+\, q \, y_1y'_1} \,\,\,  e^{- \frac{r\, (y_2^2+{y'_2}^2)}{2}\,+\, s \, y_2y'_2} \,.
\end{equation}
Writing $P(y_1,y_2;y'_1,y'_2,t)$ in terms of original variables $x_A$, $x_B$ (\ref{hamil_two_cos_ycoords}) gives
\begin{align} 
P(x_A,x_B;x'_A,x'_B=x_B,t)= \sqrt{\frac{p-q}{\pi}} \,  \sqrt{\frac{r-s}{\pi}}\,   &e^{- \frac{(p+r)}{4}\, \,(x_A^2+{x'_A}^2)\,+\, \frac{(q+s)}{2} \, x_A\, x'_A } \,  \nonumber \\
 &e^{ -\, \frac{x_B^2}{2} \,(p+r-q-s) \,  +\, x_B \, \frac{(x_A+x'_A)  }{2} \,\,(-p-s+q+r)   } \ . \label{Normalised_time_evol_dm_operator_xcoords_with_xB=x'B}
\end{align}
We now trace over the 2nd oscillator $P_A(x_A;x'_A,t)= Tr_B[P(x_A,x_B;x'_A,x'_B,t)]$.\ For this we integrate (\ref{Normalised_time_evol_dm_operator_xcoords_with_xB=x'B}) over $x_B$,\, after performing the integration, we get  
\begin{equation} \label{Reduced_time_evol_operator_os_A}
P_A(x_A;x'_A,t)= \sqrt{\frac{\gamma-\beta}{\pi}} \, e^{- \frac{\gamma}{2} (x_A^2+{x'_A}^2)\,+ \beta \, x_A \,x'_A}    \ ,
\end{equation}
where 
\begin{align}
\gamma= \frac{p+r}{2} - \frac{1}{4} \, \frac{(p+s-q-r)^2}{p+r-q-s} \,;& \qquad\beta= \frac{q+s}{2} \,+\, \frac{1}{4} \, \frac{(p+s-q-r)^2}{p+r-q-s} \,\,, \nonumber \\
\gamma-\beta= 2\, \frac{(p-q)(r-s)}{p-q+r-s} \,;& \qquad \gamma+\beta= \frac{p+q+r+s}{2}\,\, \label{Expressions_gamma_N_beta} . 
\end{align}
The entropy associated with the reduced density matrix $P_A(x_A,x'_A,t)$ is given by $S_A=- Tr(P_A \log{P_A})$. The eigenvalues $\lambda_n$ and eigenvectors $f_n(x)$ of an operator of the form (\ref{Reduced_time_evol_operator_os_A}) are given in \cite{Srednicki:1993im}: we have $\lambda_n= (1-\zeta) \, \zeta^n $, where $\zeta= \frac{\beta}{\gamma+\alpha}$, \, $\alpha=\sqrt{\gamma^2-\beta^2}$, which gives
\begin{equation} \label{VN_entropy_subsystem_A_For_time_evol_operator}
S_A= - \log(1-\zeta) \, - \, \frac{\zeta}{1-\zeta} \log{\zeta} \ .
\end{equation}
We see that the entropy $S_A$ is complex valued, recasting $\zeta$ in terms of $\gamma+\beta$ and $\gamma-\beta$,
\begin{equation} \label{Expression_zeta_}
\zeta= \frac{\sqrt{\gamma+\beta} \,-\, \sqrt{\gamma -\beta}} { \sqrt{\gamma+\beta} \,+\, \sqrt{\gamma -\beta} } \ .
\end{equation}
The explicit expressions for (\ref{Expressions_gamma_N_beta}) in terms of original variables are given by
\begin{align}
\sqrt{\gamma+ \beta} =   \Big( -i \Big( \frac{\Omega_1}{2} \cot{\frac{\Omega_1 t}{2}} + \frac{\Omega_2}{2} \cot\frac{\Omega_2 t}{2} \Big)  \Big)^{\frac{1}{2}} ,
  \quad 
\sqrt{\gamma - \beta} =    \Big( \frac{2 i} {\frac{1} {\Omega_1} \cot{\frac{\Omega_1 t}{2}} + \frac{1}{\Omega_2} \cot{ \frac{\Omega_2 t}{2}} } \Big)^{\frac{1}{2}} .   \label{Expressions_gamma_plus_minus_beta}
\end{align} 
For $\Omega_1 =\Omega_2=\omega$ (\ie\ $k_2=0$), we recover our result for two uncoupled oscillators.\,   
Comparing our result with the spacelike entanglement evaluated at finite
inverse temperature $it$, we recover the result in \cite{Katsinis:2019lis}\
(in particular $\zeta$ in (\ref{Expression_zeta_}) matches with
eq.(2.22) in \cite{Katsinis:2019lis}).


\begin{thebibliography}{} 

\footnotesize{

\bibitem{Ryu:2006bv} 
  S.~Ryu and T.~Takayanagi,
  ``Holographic derivation of entanglement entropy from AdS/CFT,''
  Phys.\ Rev.\ Lett.\  {\bf 96}, 181602 (2006)
  [hep-th/0603001].

\bibitem{Ryu:2006ef} 
  S.~Ryu and T.~Takayanagi,
  ``Aspects of Holographic Entanglement Entropy,''
  JHEP {\bf 0608}, 045 (2006)
  [hep-th/0605073].

\bibitem{HRT} 
V.~E.~Hubeny, M.~Rangamani and T.~Takayanagi,
``A Covariant holographic entanglement entropy proposal,'' 
JHEP {\bf 0707} (2007) 062  [arXiv:0705.0016 [hep-th]].

\bibitem{Maldacena:1997re}
  J.~M.~Maldacena,
  ``The large N limit of superconformal field theories and supergravity,''
  Adv.\ Theor.\ Math.\ Phys.\  {\bf 2}, 231 (1998)
  [Int.\ J.\ Theor.\ Phys.\  {\bf 38}, 1113 (1999)]
  [arXiv:hep-th/9711200].

\bibitem{Gubser:1998bc}
  S.~S.~Gubser, I.~R.~Klebanov and A.~M.~Polyakov,
  ``Gauge theory correlators from non-critical string theory,''
  Phys.\ Lett.\  B {\bf 428}, 105 (1998)
  [arXiv:hep-th/9802109].

\bibitem{Witten:1998qj}
  E.~Witten,
  ``Anti-de Sitter space and holography,''
  Adv.\ Theor.\ Math.\ Phys.\  {\bf 2}, 253 (1998)
  [arXiv:hep-th/9802150].

\bibitem{Strominger:2001pn} 
  A.~Strominger,
  ``The dS / CFT correspondence,''
  JHEP {\bf 0110}, 034 (2001)
  [hep-th/0106113].

\bibitem{Witten:2001kn} 
  E.~Witten,
  ``Quantum gravity in de Sitter space,''
  [hep-th/0106109].

\bibitem{Maldacena:2002vr}
  J.~M.~Maldacena,
  ``Non-Gaussian features of primordial fluctuations in single field inflationary models,''
  JHEP {\bf 0305}, 013 (2003),\ 
  [astro-ph/0210603].

\bibitem{Anninos:2011ui} 
  D.~Anninos, T.~Hartman and A.~Strominger,
  ``Higher Spin Realization of the dS/CFT Correspondence,''
  Class. Quant. Grav. {\bf 34} no.1, 015009 (2017)
  doi:10.1088/1361-6382/34/1/015009
  [arXiv:1108.5735 [hep-th]].

\bibitem{Doi:2022iyj}
K.~Doi, J.~Harper, A.~Mollabashi, T.~Takayanagi and Y.~Taki,
``Pseudo Entropy in dS/CFT and Time-like Entanglement Entropy,''
[arXiv:2210.09457 [hep-th]].

\bibitem{Narayan:2022afv}
K.~Narayan,
``de Sitter space, extremal surfaces, and time entanglement,''
Phys. Rev. D \textbf{107}, no.12, 126004 (2023)
doi:10.1103/PhysRevD.107.126004
[arXiv:2210.12963 [hep-th]].

\bibitem{Narayan:2015vda} 
  K.~Narayan,
  ``de Sitter extremal surfaces,''
  Phys.\ Rev.\ D {\bf 91}, no. 12, 126011 (2015)
  [arXiv:1501.03019 [hep-th]].

\bibitem{Narayan:2015oka} 
  K.~Narayan,
  ``de Sitter space and extremal surfaces for spheres,''
  Phys.\ Lett.\ B {\bf 753}, 308 (2016)
  [arXiv:1504.07430 [hep-th]].

\bibitem{Sato:2015tta} 
  Y.~Sato,
  ``Comments on Entanglement Entropy in the dS/CFT Correspondence,''
  Phys.\ Rev.\ D {\bf 91}, no. 8, 086009 (2015)
  [arXiv:1501.04903 [hep-th]].

\bibitem{Miyaji:2015yva} 
  M.~Miyaji and T.~Takayanagi,
  ``Surface/State Correspondence as a Generalized Holography,''
  PTEP {\bf 2015}, no. 7, 073B03 (2015)
  doi:10.1093/ptep/ptv089
  [arXiv:1503.03542 [hep-th]].

 \bibitem{Narayan:2017xca} 
 K.~Narayan,
 ``On extremal surfaces and de Sitter entropy,''
 Phys.\ Lett.\ B {\bf 779}, 214 (2018)
 [arXiv:1711.01107 [hep-th]].

\bibitem{Narayan:2020nsc}
K.~Narayan,
``de Sitter future-past extremal surfaces and the entanglement wedge,''
Phys. Rev. D \textbf{101}, no.8, 086014 (2020)
doi:10.1103/PhysRevD.101.086014
[arXiv:2002.11950 [hep-th]].

\bibitem{Hikida:2022ltr}
Y.~Hikida, T.~Nishioka, T.~Takayanagi and Y.~Taki,
``CFT duals of three-dimensional de Sitter gravity,''
JHEP \textbf{05}, 129 (2022)
doi:10.1007/JHEP05(2022)129
[arXiv:2203.02852 [hep-th]].

\bibitem{Hikida:2021ese}
Y.~Hikida, T.~Nishioka, T.~Takayanagi and Y.~Taki,
``Holography in de Sitter Space via Chern-Simons Gauge Theory,''
Phys. Rev. Lett. \textbf{129}, no.4, 041601 (2022)
[arXiv:2110.03197 [hep-th]].

\bibitem{Arias:2019pzy} 
  C.~Arias, F.~Diaz and P.~Sundell,
  ``De Sitter Space and Entanglement,''
  Class.\ Quant.\ Grav.\  {\bf 37}, no. 1, 015009 (2020)
  doi:10.1088/1361-6382/ab5b78
  [arXiv:1901.04554 [hep-th]].

\bibitem{Cotler:2023xku}
J.~Cotler and A.~Strominger,
``Cosmic ER=EPR in dS/CFT,''
[arXiv:2302.00632 [hep-th]].

\bibitem{Nakata:2020luh}
Y.~Nakata, T.~Takayanagi, Y.~Taki, K.~Tamaoka and Z.~Wei,
``New holographic generalization of entanglement entropy,''
Phys. Rev. D \textbf{103}, no.2, 026005 (2021)
[arXiv:2005.13801 [hep-th]].

\bibitem{Mollabashi:2020yie}
A.~Mollabashi, N.~Shiba, T.~Takayanagi, K.~Tamaoka and Z.~Wei,
``Pseudo Entropy in Free Quantum Field Theories,''
Phys. Rev. Lett. \textbf{126}, no.8, 081601 (2021)
doi:10.1103/PhysRevLett.126.081601
[arXiv:2011.09648 [hep-th]].

\bibitem{Mollabashi:2021xsd}
A.~Mollabashi, N.~Shiba, T.~Takayanagi, K.~Tamaoka and Z.~Wei,
``Aspects of pseudoentropy in field theories,''
Phys. Rev. Res. \textbf{3}, no.3, 033254 (2021)
doi:10.1103/PhysRevResearch.3.033254
[arXiv:2106.03118 [hep-th]].

\bibitem{Nishioka:2021cxe}
T.~Nishioka, T.~Takayanagi and Y.~Taki,
``Topological pseudo entropy,''
JHEP \textbf{09}, 015 (2021)
doi:10.1007/JHEP09(2021)015
[arXiv:2107.01797 [hep-th]].

\bibitem{Mukherjee:2022jac}
J.~Mukherjee,
``Pseudo Entropy in U(1) gauge theory,''
JHEP \textbf{10}, 016 (2022)
doi:10.1007/JHEP10(2022)016
[arXiv:2205.08179 [hep-th]].

\bibitem{Bhattacharya:2022wlp}
A.~Bhattacharya, A.~Bhattacharyya and S.~Maulik,
``Pseudocomplexity of purification for free scalar field theories,''
Phys. Rev. D \textbf{106}, no.8, 8 (2022)
doi:10.1103/PhysRevD.106.086010
[arXiv:2209.00049 [hep-th]].

\bibitem{Guo:2022jzs}
W.~z.~Guo, S.~He and Y.~X.~Zhang,
``Constructible reality condition of pseudo entropy via pseudo-Hermiticity,''
[arXiv:2209.07308 [hep-th]].

\bibitem{Liu:2022ugc}
B.~Liu, H.~Chen and B.~Lian,
``Entanglement Entropy in Timelike Slices: a Free Fermion Study,''
[arXiv:2210.03134 [cond-mat.stat-mech]].

\bibitem{Li:2022tsv}
Z.Li, Z.Q.Xiao, R.Q.Yang,
``On holographic time-like entanglement entropy,''
arXiv:2211.14883[hep-th].

\bibitem{Doi:2023zaf}
K.~Doi, J.~Harper, A.~Mollabashi, T.~Takayanagi and Y.~Taki,
``Timelike entanglement entropy,''
[arXiv:2302.11695 [hep-th]].

\bibitem{Jiang:2023ffu}
X.~Jiang, P.~Wang, H.~Wu and H.~Yang,
``Timelike entanglement entropy and $T\bar{T}$ deformation,''
[arXiv:2302.13872 [hep-th]].

\bibitem{Chen:2023gnh}
Z.~Chen,
``Complex-valued Holographic Pseudo Entropy via Real-time AdS/CFT Correspondence,''
[arXiv:2302.14303 [hep-th]].

\bibitem{Wang:2018jva}
P.~Wang, H.~Wu and H.~Yang,
``Fix the dual geometries of $T\bar{T}$ deformed CFT$_2$ and highly excited states of CFT$_2$,''
Eur. Phys. J. C \textbf{80}, no.12, 1117 (2020)
doi:10.1140/epjc/s10052-020-08680-7
[arXiv:1811.07758 [hep-th]].

\bibitem{Holzhey:1994we} 
  C.~Holzhey, F.~Larsen and F.~Wilczek,
  ``Geometric and renormalized entropy in conformal field theory,''
  Nucl.\ Phys.\ B {\bf 424}, 443 (1994)
  [hep-th/9403108].

\bibitem{Calabrese:2004eu} 
  P.~Calabrese and J.~L.~Cardy,
  ``Entanglement entropy and quantum field theory,''
  J.\ Stat.\ Mech.\  {\bf 0406}, P06002 (2004)
  [hep-th/0405152].

\bibitem{Calabrese:2009qy} 
  P.~Calabrese and J.~Cardy,
  ``Entanglement entropy and conformal field theory,''
  J.\ Phys.\ A {\bf 42}, 504005 (2009)
  doi:10.1088/1751-8113/42/50/504005
  [arXiv:0905.4013 [cond-mat.stat-mech]].

\bibitem{Chen:2020tes}
Y.~Chen, V.~Gorbenko and J.~Maldacena,
``Bra-ket wormholes in gravitationally prepared states,''
JHEP \textbf{02}, 009 (2021)
doi:10.1007/JHEP02(2021)009
[arXiv:2007.16091 [hep-th]].

\bibitem{Goswami:2021ksw}
K.~Goswami, K.~Narayan and H.~K.~Saini,
``Cosmologies, singularities and quantum extremal surfaces,''
JHEP \textbf{03}, 201 (2022)
doi:10.1007/JHEP03(2022)201
[arXiv:2111.14906 [hep-th]].

\bibitem{Sakurai}
J. Sakurai, ``Modern Quantum Mechanics'', Revised edn.

\bibitem{Jonay:2018yei}
C.~Jonay, D.~A.~Huse and A.~Nahum,
``Coarse-grained dynamics of operator and state entanglement,''
[arXiv:1803.00089 [cond-mat.stat-mech]].

\bibitem{Castellani:2021zrk}
L.~Castellani,
``Entropy of temporal entanglement,''
[arXiv:2104.05722 [quant-ph]].

\bibitem{Lerose:2021svg}
A.~Lerose, M.~Sonner and D.~A.~Abanin,
``Scaling of temporal entanglement in proximity to integrability,''
Phys. Rev. B \textbf{104}, no.3, 035137 (2021)
doi:10.1103/PhysRevB.104.035137
[arXiv:2104.07607 [quant-ph]].

\bibitem{Diaz:2021snw}
N.~L.~Diaz, J.~M.~Matera and R.~Rossignoli,
``Path Integrals from Spacetime Quantum Actions,''
[arXiv:2111.05383 [quant-ph]].

\bibitem{dresselWeak}
  J. Dressel, M. Malik, F. M. Miatto, A. N. Jordan, R. W. Boyd,
  ``Colloquium: Under- standing Quantum Weak Values: Basics and Applications,''
  Rev. Mod. Phys. 86 (2014) 307 [arXiv:1305.7154 [quant-ph]].

\bibitem{SalekPostselection}
  S. Salek, R. Schubert, and K. Wiesner,
  ``Negative conditional entropy of postselected states,''
  Phys. Rev. A 90 (2014) 022116 [arXiv:1305.0932 [quant-ph]].

\bibitem{Narayan:2023zen}
K.~Narayan,
``Further remarks on de Sitter space, extremal surfaces, and time entanglement,''
Phys. Rev. D \textbf{109}, no.8, 086009 (2024)
doi:10.1103/PhysRevD.109.086009
[arXiv:2310.00320 [hep-th]].

\bibitem{Hayden:2011ag}
P.~Hayden, M.~Headrick and A.~Maloney,
``Holographic Mutual Information is Monogamous,''
Phys. Rev. D \textbf{87}, no.4, 046003 (2013)
doi:10.1103/PhysRevD.87.046003
[arXiv:1107.2940 [hep-th]].

\bibitem{Jatkar:2017jwz} 
  D.~P.~Jatkar and K.~Narayan,
  ``Ghost-spin chains, entanglement and $bc$-ghost CFTs,''
  Phys.\ Rev.\ D {\bf 96}, no. 10, 106015 (2017)
  [arXiv:1706.06828 [hep-th]].

\bibitem{Narayan:2016xwq} 
  K.~Narayan,
  ``On $dS_4$ extremal surfaces and entanglement entropy in some ghost CFTs,''
  Phys.\ Rev.\ D {\bf 94}, no. 4, 046001 (2016)
  [arXiv:1602.06505 [hep-th]].

\bibitem{merzbacher}
  E. Merzbacher, ``Quantum Mechanics'', 3rd edn.

\bibitem{Srednicki:1993im}
M.~Srednicki,
``Entropy and area'',
''Phys. Rev. Lett. \textbf{71}, 666-669 (1993)
doi:10.1103/PhysRevLett.71.666
[arXiv:hep-th/9303048 [hep-th]].

\bibitem{Katsinis:2019lis}
Katsinis, Dimitrios and Pastras, Georgios,
``An Inverse Mass Expansion for the Mutual Information in Free Scalar QFT at Finite Temperature'',
JHEP \textbf{02}, 091 (2020)
doi:10.1007/JHEP02(2020)091
[arXiv:1907.08508 [hep-th]].


}
\end{thebibliography}
\end{document}